# Zero-DeepSub: Zero-Shot Deep Subspace Reconstruction for Rapid Multiparametric Quantitative MRI Using 3D-QALAS

Yohan Jun[1,2], Yamin Arefeen[3,4], Jaejin Cho[1,2], Shohei Fujita[1,2], Xiaoqing Wang[1,2], P. Ellen Grant[2,5], Borjan Gagoski[2,5], Camilo Jaimes[2,6], Michael S. Gee[2,6,*], and Berkin Bilgic[1,2,7,*]

[1]Athinoula A. Martinos Center for Biomedical Imaging, Massachusetts General Hospital, Boston, MA, United States

[2]Department of Radiology, Harvard Medical School, Boston, MA, United States

[3]Chandra Family Department of Electrical and Computer Engineering, The University of Texas, Austin, TX, United States

[4]Department of Imaging Physics, The University of Texas MD Anderson Cancer Center, Houston, TX, United States

[5]Fetal-Neonatal Neuroimaging & Developmental Science Center, Boston Children's Hospital, Boston, MA, United States

[6]Department of Radiology, Massachusetts General Hospital, Boston, MA, United States

[7]Harvard/MIT Health Sciences and Technology, Cambridge, MA, United States

* Equal contribution as last authors

The word count of the body of the text: ~5000

Corresponding Author: Yohan Jun

Athinoula A. Martinos Center for Biomedical Imaging, Building 149, 13th Street, Rm 2301, Charlestown, MA 02129, USA.

Email: yjun@mgh.harvard.edu

Grant Support: This work was supported by research grants NIH R01 EB032378, R01 EB028797, R03 EB031175, P41 EB030006, U01 EB026996, U01 DA055353, UG3 EB034875 and the NVIDIA Corporation for computing support.

Running Title: Zero-Shot Deep Subspace Reconstruction for Rapid Multiparametric Quantitative MRI Using 3D-QALAS

**Abstract**

**Purpose:** To develop and evaluate methods for 1) reconstructing 3D-quantification using an interleaved Look-Locker acquisition sequence with $T_2$ preparation pulse (3D-QALAS) time-series images using a low-rank subspace method, which enables accurate and rapid $T_1$ and $T_2$ mapping, and 2) improving the fidelity of subspace QALAS by combining scan-specific deep-learning-based reconstruction and subspace modeling.

**Methods:** A low-rank subspace method for 3D-QALAS (i.e., subspace QALAS) and zero-shot deep-learning subspace method (i.e., Zero-DeepSub) were proposed for rapid and high fidelity $T_1$ and $T_2$ mapping and time-resolved imaging using 3D-QALAS. Using an ISMRM/NIST system phantom, the accuracy and reproducibility of the $T_1$ and $T_2$ maps estimated using the proposed methods were evaluated by comparing them with reference techniques. The reconstruction performance of the proposed subspace QALAS using Zero-DeepSub was evaluated *in vivo* and compared with conventional QALAS at high reduction factors of up to 9-fold.

**Results:** Phantom experiments showed that subspace QALAS had good linearity with respect to the reference methods while reducing biases and improving precision compared to conventional QALAS, especially for $T_2$ maps. Moreover, *in vivo* results demonstrated that subspace QALAS had better g-factor maps and could reduce voxel blurring, noise, and artifacts compared to conventional QALAS and showed robust performance at up to 9-fold acceleration with Zero-DeepSub, which enabled whole-brain $T_1$, $T_2$, and PD mapping at 1 mm isotropic resolution within 2 min of scan time.

**Conclusion:** The proposed subspace QALAS along with Zero-DeepSub enabled high fidelity and rapid whole-brain multiparametric quantification and time-resolved imaging.

# INTRODUCTION

Quantitative magnetic resonance imaging (qMRI) (1,2) has been utilized for neuroimaging studies and diagnosis of diseases, such as Alzheimer's disease (3,4), multiple sclerosis (5–7), brain tumors (8,9), memory loss (10), and aging analyses (11,12), since it provides quantitative information on human tissue characteristics, such as $T_1$, $T_2$, and $T_2^*$ relaxation rates, and magnetic susceptibility. Quantitative magnetic resonance (MR) parameter maps can be obtained separately using inversion-recovery spin-echo (IR-SE) or fast-spin-echo (IR-FSE) for $T_1$ and single-echo spin-echo (SE-SE) or fast-spin-echo (SE-FSE) for $T_2$ maps, which are usually used for gold standard methods. Alternative approaches, such as DESPOT1, DESPOT2, MP2RAGE, MPnRAGE, MOLLI, and GRASE, also have been used for rapid mapping of a single relaxation parameter (13–17). However, multiple sequences need to be run to acquire other maps of interest, prolonging the total scan time and increasing vulnerability to motion between the sequences.

Numerous MRI techniques have been proposed for simultaneously acquiring multiparametric maps by designing an advanced pulse sequence, including magnetic resonance fingerprinting (MRF) (18), 3D-quantification using an interleaved Look-Locker acquisition sequence with $T_2$ preparation pulse (3D-QALAS) (19), MR multitasking (20), echo planar time-resolved imaging (EPTI) (21), and quantification of relaxation times and proton density by the multi-echo acquisition of a saturation-recovery using turbo-spin-echo readout (QRAPMASTER) (22).

In particular, 3D-QALAS, initially proposed for cardiac imaging (19,23), enabled whole-brain $T_1$, $T_2$, and proton density (PD) mapping at 1 mm isotropic resolution within 11 min (24,25) and yielded highly repeatable $T_1$, $T_2$, and PD values and high scan-rescan repeatability of subcortical brain volumes and cortical thickness measurements (26,27). 3D-QALAS also estimates $T_1$ and $T_2$ values with strong linearity with respect to reference values (27). There have been attempts to reduce the acquisition time by employing compressed sensing (28) and wave-controlled aliasing in parallel imaging (Wave-CAIPI) (29), which uses generalized parallel imaging reconstructions along with corkscrew k-space trajectories enabling 6-fold acceleration (30). A deep-learning-based method called Wave-MoDL further accelerated 3D-QALAS up to 12-fold by combining Wave-CAIPI and deep model-based reconstruction (25). This supervised deep-learning strategy required an external training database with high-quality and fully-sampled data.

3D-QALAS has five fast low-angle shot (FLASH) readouts within each QALAS block to generate high-resolution $T_1$, $T_2$, and PD maps. These parameter maps can be estimated by matching the acquired images to a Bloch-simulated dictionary. However, the fitting process inherently assumes that the entire k-space for each acquisition is acquired instantly at the first echo that samples the center of the k-space. This assumption neglects $T_1$ relaxation during the lengthy echo train of the FLASH readout, which might cause blurring and biases (31) in the reconstructed $T_1$ and $T_2$ maps. Thus, time-resolved image reconstruction and fitting, which utilizes the full QALAS signal evolution, including the signal changes during the entire echo train length (ETL), could yield more accurate $T_1$ and $T_2$ estimation in place of the conventional five-point dictionary that does not account for signal relaxation during the acquisition.

Low-rank subspace and shuffling methods have emerged as powerful methods for reconstructing time-resolved MR images and qMRI since they incorporate low-rank subspace bases that are calculated from Bloch equations (20,21,32–36). $T_2$-shuffling, for instance, showed multi-contrast and sharp $T_2$-weighted images by leveraging the $T_2$ relaxation during the fast spin echo (FSE) readout using a low-rank subspace method and shuffled k-space data acquisition (33). 3D-EPTI acquires highly undersampled k-t data using an inversion-recovery gradient-echo (IR-GE) and a variable flip angle gradient and spin-echo (VFA-GRASE) and also exploits a low-rank subspace approach to reconstruct time-series data efficiently (37).

In this study, we propose reconstructing QALAS time-series data using a low-rank subspace method (i.e., ***subspace QALAS***), thus enabling accurate $T_1$ and $T_2$ mapping with reduced biases, g-factor noise amplification, and relaxation-related blurring compared to conventional QALAS. To the best of our knowledge, this is the first study to reconstruct QALAS time-series images and obtain $T_1$, $T_2$, and PD maps using a low-rank subspace method, which utilizes the full QALAS signal evolution.

Besides subspace modeling, machine- and deep-learning-based techniques have found increasing applications in MRI as well. In MR image reconstruction, a combination of a deep-learning-based regularizer and parallel imaging forward model outperformed conventional parallel imaging and compressed sensing (38–43). Deep-learning-based methods have also been applied to subspace reconstruction problems (44–47). For example, a deep subspace learning method combined a deep network with a low-rank subspace modeling, thus allowing improved performance in $T_1$ mapping using a single-shot IR radial FLASH sequence (46). However, those methods based on a supervised training paradigm demand fully sampled or reference data for training deep networks, which hinders their application in qMRI reconstruction problems where it may be prohibitively difficult to obtain fully sampled or reference data. To tackle this problem, self-supervised or unsupervised learning paradigms, which do not require labeled data for model training, have gained attention in recent years and have been applied to MRI denoising, segmentation, reconstruction, and qMRI (48–52). In particular, self-supervised learning via data undersampling (SSDU) and zero-shot self-supervised learning (ZS-SSL), which do not require fully sampled k-space data, were proposed for MR image reconstruction and demonstrated comparable reconstruction performance with supervised learning-based methods (50,53).

Here, inspired by these previous works, we propose a zero-shot deep-learning subspace method, ***Zero-DeepSub***, which combines scan-specific deep-learning-based reconstruction with low-rank subspace modeling, to further improve the fidelity of multiparametric qMRI, specifically for our proposed subspace QALAS acquisition. A deep model-based architecture is designed for reconstructing subspace coefficients from the acquired multi-echo k-space data where the deep network can be trained only using the acquired undersampled k-space data in a scan-specific manner to denoise the subspace coefficients without fully sampled k-space or external datasets. Our main contributions are as follows:

- We propose "***subspace QALAS***" to reconstruct QALAS time-series images using a low-rank subspace method, thus enabling more accurate $T_1$ and $T_2$ mapping with reduced blurring and g-factor noise amplification compared to conventional QALAS.
- We develop "***Zero-DeepSub***" to further improve the fidelity of subspace QALAS by combining a scan-specific deep-learning-based reconstruction and subspace modeling, which reconstructs denoised subspace coefficients from the acquired multi-echo k-space that can be used for generating quantitative maps.
- Our proposed methods were validated on an International Society for Magnetic Resonance in Medicine and National Institute of Standards and Technology (ISMRM/NIST) system phantom, where subspace QALAS demonstrated good linearity with respect to the gold standard methods, including IR-FSE and SE-FSE, and had reduced biases compared to conventional QALAS, especially for $T_2$ maps.
- *In vivo* results demonstrated that subspace QALAS showed better g-factor maps with reduced voxel blurring and noise compared to the conventional QALAS and could accelerate 3D-QALAS scans by up to 9-fold using Zero-DeepSub, thus enabling whole-brain $T_1$, $T_2$, and PD mapping at 1 mm isotropic resolution within 2 min.
- All the source codes can be found here: https://github.com/yohan-jun/Zero-DeepSub

**THEORY**

**QALAS Signal Model**

The MR signal model describing acquired k-space data with multichannel receive arrays can be represented as follows:

$$y_l(t) = \int_\mathbf{r} c_l(\mathbf{r}) S_t(\mathbf{r}) e^{-i\mathbf{r}2\pi\mathbf{k}(t)} d\mathbf{r}, \qquad [1]$$

where $y_l(t)$ is the obtained MR signal from the $l$-th coil at the time $t$ ($0 \leq t < \text{TR}$), TR is the duration of QALAS block, $c_l(\mathbf{r})$ is the $l$-th ($l = 1, \ldots, L$) coil sensitivity at the spatial location $\mathbf{r}$, $S_t(\mathbf{r})$ is the transverse magnetization, and $\mathbf{k}(t)$ is the k-space sampling trajectory.

Referring to the original QALAS paper (19), the longitudinal magnetization of QALAS can be represented in terms of $t$. The sequence diagram of QALAS is presented in Fig. 1a. There are five FLASH readouts within each QALAS block, and $T_1$ relaxation occurs between the acquisitions during the time interval $\Delta t$, which can be described as follows:

$$M_{t+\Delta t}(\mathbf{r}) = M_0(\mathbf{r}) - \big(M_0(\mathbf{r}) - M_t(\mathbf{r})\big) e^{-\frac{\Delta t}{T_1(\mathbf{r})}}, \qquad [2]$$

with

$$t = [\text{T}_{c-d}, \text{T}_{e-f}, \text{T}_{g-h}, \text{T}_{i-j}, \text{T}_{k-l}, \text{T}_{m-a}], \qquad [3]$$

where $t = \text{T}_{(\cdot)}$ is the time of magnetization of $\text{M}_{(\cdot)}$ and $t = \text{T}_{p-q}$ is the time points between $t = \text{T}_p$ and $t = \text{T}_q$.

While acquiring the signal for five FLASH readouts (54), the longitudinal magnetization follows $T_1^*$ relaxation with the saturated longitudinal signal $M_0^*$ based on the Ernst equation (55,56) during the time interval $\Delta t$, where $\Delta t = \text{T}_{\text{ES}}$ (echo spacing), which can be represented as follows:

$$M_{t+\Delta t}(\mathbf{r}) = M_0^*(\mathbf{r}) - \big(M_0^*(\mathbf{r}) - M_t(\mathbf{r})\big) e^{-\frac{\Delta t}{T_1^*(\mathbf{r})}}, \qquad [4]$$

with

$$t = [T_{b-c}, T_{f-g}, T_{h-i}, T_{j-k}, T_{l-m}], \quad [5]$$

$$M_0^*(\mathbf{r}) = \frac{1-e^{-\frac{T_{ES}}{T_1(\mathbf{r})}}}{1-\cos(\alpha)e^{-\frac{T_{ES}}{T_1(\mathbf{r})}}} M_0(\mathbf{r}), \quad [6]$$

$$T_1^*(\mathbf{r}) = \frac{1-e^{-\frac{T_{ES}}{T_1(\mathbf{r})}}}{1-\cos(\alpha)e^{-\frac{T_{ES}}{T_1(\mathbf{r})}}} T_1(\mathbf{r}), \quad [7]$$

where $\alpha$ is the flip angle. Then, the acquired transverse magnetization signals can be represented as follows:

$$S_{t+\Delta t}(\mathbf{r}) = \left\{ M_0^*(\mathbf{r}) - \left(M_0^*(\mathbf{r}) - M_t(\mathbf{r})\right) e^{-\frac{\Delta t}{T_1^*(\mathbf{r})}} \right\} \cdot \sin(\alpha). \quad [8]$$

Considering the $B_1^+$ inhomogeneity, the flip angle $\alpha$ is dependent on the spatial location $\mathbf{r}$ and can be represented as follows: $\alpha(\mathbf{r}) = \alpha \cdot B_1^+(\mathbf{r})$.

During the $T_2$-sensitizing time using an adiabatic $T_2$ preparation pulse, $T_2$ relaxation occurs between the time points $T_{a-b}$, which can be described as follows:

$$M_{t+TE_{T2prep}}(\mathbf{r}) = M_t(\mathbf{r}) e^{-\frac{TE_{T2prep}}{T_2(\mathbf{r})}}, \quad [9]$$

where $TE_{T2prep}$ is the time interval between the 90-degree tip-down and 90-degree tip-up radiofrequency (RF) pulses.

During the $T_1$-sensitizing time using a 180-degree inversion pulse, the longitudinal signal is inverted between the time points $T_{d-e}$, which can be described as follows:

$$M_t(\mathbf{r}) = -M_t(\mathbf{r}) \cdot IE(\mathbf{r}), \quad [10]$$

where IE is the inversion efficiency of the inversion pulse, which ranges between 0 (i.e., no inversion) to 1 (i.e., perfect inversion).

**Subspace QALAS**

In the conventional QALAS reconstruction, $T_1$ relaxation across time $t$ during the readouts (in Eqs. 4-8) is ignored by assuming that each k-space data is acquired instantly at the first echo (i.e., the center of the k-space in a center-out acquisition) of the lengthy ETL of FLASH readouts. The assumption might cause voxel blurring in image space and quantification bias (31). Instead, we propose utilizing signal relaxation properties during the readouts and reconstructing time-series QALAS data. The overall reconstruction scheme of the proposed subspace QALAS method is presented in Fig. 1b.

We consider that $T$ echo images are acquired in each QALAS block where the number of echoes is $T = \text{ETL} \times 5$ (i.e., ETL = echo train length of each FLASH readout, 5 = the number of FLASH readouts in each QALAS block) where $T > 600$ in practice. Similar to the shuffling method (33), the $i$-th echo image ($i = 1, ..., T$) with $T_{ES}$ echo spacing can be described as follows:

$$E_i(\mathbf{r}) = S_{f(i)}(\mathbf{r}), \quad [11]$$

with

$$f(i) = T_i + T_{ES}\left[\frac{i}{\text{ETL}}\right], \quad [12]$$

where $T_i$ is the time when the readout starts (i.e., $T_i = T_b, T_f, T_h, T_j,$ and $T_l$ for each readout), and the multi-echo images can be represented as follows:

$$\mathbf{E}(\mathbf{r}) = [E_1(\mathbf{r})\ E_2(\mathbf{r})\ \cdots\ E_T(\mathbf{r})]^\mathsf{T}, \quad [13]$$

where $\mathbf{E} \in \mathbb{C}^{NT}$ is a vector. Though the temporal dimension ($T$) of $\mathbf{E}$ is large (e.g., $T > 600$ in practice), $\mathbf{E}$ can be represented using a low-dimensional subspace (33) since the signal evolution of QALAS time-series data is highly correlated in the temporal dimension:

$$\mathbf{E} = \mathbf{\Phi}\mathbf{\Phi}^H\mathbf{E} \approx \mathbf{\Phi}_K\mathbf{\Phi}_K^H\mathbf{E}$$

$$s.t. \quad \|\mathbf{E} - \mathbf{\Phi}_K\mathbf{\Phi}_K^H\mathbf{E}\| < \epsilon, \quad [14]$$

where $\mathbf{\Phi}$ is an orthonormal basis of the QALAS signal evolution, $\mathbf{\Phi}_K$ is a $K$-dimension ($K \ll T$) subspace basis, and $\epsilon$ is the error criteria where the error can be calculated between the signal evolution generated based on the QALAS signal model and the generated signal evolution using the subspace basis $\mathbf{\Phi}_K$. Instead of reconstructing the whole high-dimensional multi-echo images, low-dimensional subspace coefficients can be reconstructed using the subspace basis, which is:

$$\mathbf{x} = \mathbf{\Phi}_K^H\mathbf{E}, \quad [15]$$

where $\mathbf{x} \in \mathbb{C}^{NK}$ are the subspace coefficients to be reconstructed. After reconstruction of the subspace coefficients, multi-echo images $\mathbf{E} \in \mathbb{C}^{NT}$ can be obtained using the subspace basis $\mathbf{\Phi}_K$:

$$\mathbf{E} = \mathbf{\Phi}_K\mathbf{x}. \quad [16]$$

Then, a low-rank subspace-based reconstruction problem can be formulated as the following equation:

$$\min_{\mathbf{x}} \|\mathbf{y} - \mathbf{A}\mathbf{x}\|_2^2 + \lambda \mathcal{R}(\mathbf{x}), \qquad [17]$$

with the forward operator $\mathbf{A}$:

$$\mathbf{A} = \mathbf{MFC}\mathbf{\Phi}_K : \mathbb{C}^{NK} \to \mathbb{C}^{NTL}, \qquad [18]$$

where $\mathbf{y} \in \mathbb{C}^{NTL}$ is the acquired multi-echo and multi-coil k-space data, which has been decomposed and zero-padded across the ETL index ($T = \text{ETL} \times 5$) on $k_y - k_z$ plane from the original acquired multi-coil k-space data. $\mathbf{x} \in \mathbb{C}^{NK}$ are the desired subspace coefficient images, $\mathbf{A}$ is the forward operator, which transforms the subspace coefficients into multi-echo and multi-coil k-space data, containing a k-space sampling mask $\mathbf{M} \in \mathbb{R}^{NTL \times NTL}$, Fourier transform $\mathbf{F} \in \mathbb{C}^{NTL \times NTL}$, coil sensitivity map $\mathbf{C} \in \mathbb{C}^{NTL \times NT}$, and subspace matrix $\mathbf{\Phi}_K \in \mathbb{R}^{NT \times NK}$, which is constructed from a subspace basis. $N$ and $L$ are the matrix size of the image and the number of coils, respectively. $\mathcal{R}$ is the regularization term and $\lambda$ is the regularization parameter that controls the balance between the data consistency and regularization terms.

*LLR and l₁-Wavelet Regularizations*

For a low-rank subspace-based reconstruction problem, locally low-rank (LLR) $\mathcal{R}_{\text{LLR}}(\cdot)$ (33,37,57) or $l_1$-wavelet-based regularization $\mathcal{R}_{\text{wavelet}}(\cdot)$ (28) can be used to solve Eq. 17 as follows:

$$\mathcal{R}_{\text{LLR}}(\mathbf{x}) = \sum_{b \in \Omega} \|Q_b \mathbf{x}\|_*, \qquad [19]$$

$$\mathcal{R}_{\text{wavelet}}(\mathbf{x}) = \|\Psi \mathbf{x}\|_1, \qquad [20]$$

where $\Omega$ is the set of the image blocks of the subspace coefficient images that are generated from the original subspace coefficient images $\mathbf{x}$, $Q_b$ is the operator that extracts one of the image blocks from the set $\Omega$ and reshapes it into a Casorati matrix, and the nuclear norm $\|\cdot\|_*$ is applied to the matrix. $\Psi$ is a wavelet sparsifying transform.

*Zero-DeepSub: Zero-Shot Deep Subspace Reconstruction*

A deep-learning-based regularizer $\mathcal{R}_{\mathrm{DL}}(\cdot)$, which has been used for MR image reconstruction (38,40,42,43), can be adapted to a low-rank subspace-based reconstruction problem, which is defined:

$$\mathcal{R}_{\mathrm{DL}}(\mathbf{x}) = \|\mathbf{x} - \mathcal{D}(\mathbf{x}; \boldsymbol{\theta})\|_2^2, \quad [21]$$

where $\mathcal{D}(\cdot; \boldsymbol{\theta})$ is a deep-learning-based denoiser with trainable parameters $\boldsymbol{\theta}$ where a convolutional neural network (CNN)-based architecture is used. Substituting the regularization term in Eq. 17 with the deep-learning-based regularization term defined in Eq. 21, yields:

$$\min_{\mathbf{x}} \|\mathbf{y} - \mathbf{A}\mathbf{x}\|_2^2 + \lambda \|\mathbf{x} - \mathcal{D}(\mathbf{x}; \boldsymbol{\theta})\|_2^2, \quad [22]$$

and the unrolled network (43) that recursively updates $\mathbf{x}$ can be represented with the normal equations:

$$\mathbf{x}_{p+1} = (\mathbf{A}^H \mathbf{A} + \lambda \mathbf{I})^{-1} \left( \mathbf{A}^H \mathbf{y} + \lambda \mathcal{D}(\mathbf{x}_p; \boldsymbol{\theta}) \right), \quad [23]$$

where $\mathbf{x}_p$ is the reconstructed subspace coefficients at iteration $p$ ($p = 1, \ldots, P$) and $P$ is the number of iterations. Here, the data consistency term $(\mathbf{A}^H \mathbf{A} + \lambda \mathbf{I})^{-1}$ can be solved using a conjugate gradient optimization algorithm. Detailed implementation of the data consistency term can be found in the *Implementation Details* section.

In a supervised learning paradigm, fully sampled k-space data are used for loss calculation to find the optimal parameters of $\mathcal{D}(\cdot; \boldsymbol{\theta})$. However, for a low-rank subspace reconstruction problem, the acquisition of fully sampled k-space data in many qMRI sequences is impractical, and a supervised learning scheme may not be feasible for model training. Especially, subspace QALAS, as proposed in the ***Subspace QALAS*** section, considers that $T$ echo images are acquired in each QALAS block by decomposing the original acquired k-space data across the ETL index ($T = \mathrm{ETL} \times 5 > 600$); thus, full k-space acquisition for more than 600 echoes in each QALAS block is impractical.

To tackle the problem, we propose a zero-shot deep-learning subspace method (i.e., ***Zero-DeepSub***) combining scan-specific deep-learning-based reconstruction with low-rank subspace modeling. The detailed architecture of Zero-DeepSub is presented in Fig. 2. The original ZS-SSL was proposed for scan-specific MR image reconstruction without using fully sampled k-space data or external datasets (50). We further extended the approach to a low-rank subspace reconstruction and developed Zero-DeepSub for reconstructing subspace coefficients from highly undersampled multi-echo k-space data for multiparametric qMRI.

Similar to the k-space sampling strategy used in (50), the decomposed multi-echo QALAS k-space data **y** across the ETL index are further split into three different k-space data for training and validation, as follows:

$$\mathbf{y}_i = \mathbf{y}_i^V + \mathbf{y}_i^W + \mathbf{y}_i^Z, \qquad [24]$$

where $\mathbf{y}_i$ is the $i$-th echo k-space data, $\mathbf{y}_i^V$, $\mathbf{y}_i^W$, and $\mathbf{y}_i^Z$ are the decomposed k-space data using k-space sampling locations $V$, $W$, and $Z$, respectively, where $V$, $W$, and $Z$ are pairwise disjoint. The set of $V$, $W$, and $Z$ is different across the echo index $i$ and can also be generated differently in every training iteration or epoch, which can facilitate data incoherence between the iterations or epochs; thus, the model can be trained effectively with various subsampled k-space data generated from a single scan. Then, the training loss $\mathcal{L}_{\mathrm{TRN}}$ and validation loss $\mathcal{L}_{\mathrm{VAL}}$ can be defined as follows:

$$\mathcal{L}_{\mathrm{TRN}} = \sum_{b=1}^{B}\sum_{i=1}^{T} \mu\left\|\mathbf{y}_i^{W_b} - \mathbf{A}_{W_b}\mathbf{x}_P^{V_b}\right\|_1^1 + (1-\mu)\left\|\mathbf{y}_i^{W_b} - \mathbf{A}_{W_b}\mathbf{x}_P^{V_b}\right\|_2^2$$

$$= \sum_{b=1}^{B}\sum_{i=1}^{T} \mu\left\|\mathbf{y}_i^{W_b} - \mathbf{M}^{W_b}\mathbf{F}\mathbf{C}\mathbf{\Phi}_K\mathbf{x}_P^{V_b}\right\|_1^1 + (1-\mu)\left\|\mathbf{y}_i^{W_b} - \mathbf{M}^{W_b}\mathbf{F}\mathbf{C}\mathbf{\Phi}_K\mathbf{x}_P^{V_b}\right\|_2^2,$$

$$\mathcal{L}_{\mathrm{VAL}} = \sum_{i=1}^{T} \mu\left\|\mathbf{y}_i^Z - \mathbf{A}_Z\mathbf{x}_P^{(V+W)}\right\|_1^1 + (1-\mu)\left\|\mathbf{y}_i^Z - \mathbf{A}_Z\mathbf{x}_P^{(V+W)}\right\|_2^2$$

$$= \sum_{i=1}^{T} \mu\left\|\mathbf{y}_i^Z - \mathbf{M}^Z\mathbf{F}\mathbf{C}\mathbf{\Phi}_K\mathbf{x}_P^{(V+W)}\right\|_1^1 + (1-\mu)\left\|\mathbf{y}_i^Z - \mathbf{M}^Z\mathbf{F}\mathbf{C}\mathbf{\Phi}_K\mathbf{x}_P^{(V+W)}\right\|_2^2, \qquad [25]$$

where $B$ is the number of the generated set of $V_b$, $W_b$, and $Z_b$. $\mathbf{M}^{(\cdot)}$ is the sampling mask and $\mathbf{x}_P^{(\cdot)}$ is the reconstructed subspace coefficients using given k-space sampling locations. $\mu$ and $(1-\mu)$ are the regularization parameters of $l_1$ and $l_2$ norm loss, respectively. For training, trainable parameters $\boldsymbol{\theta}$ of the model are updated using the training loss $\mathcal{L}_{\mathrm{TRN}}$, where sampling location $V_b$ is used for data consistency and $W_b$ for loss calculation, respectively. For validation, updated parameters at specific iterations are used for validation loss $\mathcal{L}_{\mathrm{VAL}}$, where sampling location $(V+W)$ is used for data consistency and $Z$ for loss calculation, respectively. For inference, the original sampling mask is used for obtaining the output subspace coefficients using the trained parameters.

**METHODS**

**Implementation Details**

*Zero-DeepSub*

Network Architecture

The architecture of Zero-DeepSub is designed based on a deep model-based framework (38,43,58). The initial subspace coefficient images calculated by matrix multiplication with the subspace basis were concatenated along the channel dimension and fed into the deep model-based network. The CNN-based denoiser is based on residual learning, which has five CNN blocks where each block consists of a 3 × 3 convolutional layer with 128 feature maps, batch normalization layer, and leaky rectified linear unit (ReLU) activation function with 0.05 negative slope coefficient. The trainable parameter $\lambda$ was initialized as 0.005. The trainable parameters in the CNN-based denoiser were shared across iterations. $\mu$ was set as 0.5, for the regularization parameters of $l_1$ and $l_2$ norm loss. The model was trained with Adam optimizer with $\beta_1 = 0.9$ and $\beta_2 = 0.999$ with a learning rate of 0.0005 and implemented using the TensorFlow library (59). The training of the model took about 4 h for a multi-slice whole-brain reconstruction using a single NVIDIA Tesla V100 GPU.

Data Consistency Layer

The data consistency layer implemented with a conjugate gradient algorithm has high computational cost since it requires the repeated computation of forward and adjoint operations of **A**:

$$\mathbf{A}^H \mathbf{A} \mathbf{x} = \mathbf{\Phi}_K^H \mathbf{C}^H \mathbf{F}^H \mathbf{M} \mathbf{F} \mathbf{C} \mathbf{\Phi}_K \mathbf{x}. \qquad [26]$$

For subspace QALAS, it at least requires $T$ (the number of echoes) $\times$ $L$ (the number of coils) $\approx 5000$ Fourier transforms along with $T \times K$ (the number of subspace basis) $\approx 2500$ matrix multiplications for each voxel, which could not fit into GPU memory. Using a similar approach to the shuffling method (33), where the matrix multiplication order of Fourier transform and subspace basis can be changed as $\mathbf{F} \mathbf{C} \mathbf{\Phi}_K = \mathbf{\Phi}_K \mathbf{F} \mathbf{C}$, the normal equation can be rewritten as follows:

$$\mathbf{A}^H \mathbf{A} \mathbf{x} = \mathbf{C}^H \mathbf{F}^H \mathbf{\Phi}_K^H \mathbf{M} \mathbf{\Phi}_K \mathbf{F} \mathbf{C} \mathbf{x}, \qquad [27]$$

where the matrix size of **C**, **F**, and **M** can be changed: $\mathbf{C} \in \mathbb{C}^{NKL \times NK}$, $\mathbf{F} \in \mathbb{C}^{NKL \times NK}$, and $\mathbf{M} \in \mathbb{R}^{NT \times NT}$. The computation of Fourier transform is reduced from $T \times L \approx 5000$ to $K \times L \approx 30$. Here, $\mathbf{\Phi}_K^H \mathbf{M} \mathbf{\Phi}_K \in \mathbb{C}^{NK \times NK}$ is a spatio-temporal kernel that can be precomputed using subspace basis and k-space sampling mask (33), which can replace the computation of matrix multiplication of $T \times K \approx 2500$ to $K \times K = 16$ Hadamard product (i.e., element-wise multiplication), thus further reducing the computation in a data consistency layer.

k-space Sampling Scheme for Training and Validation

For the k-space sampling strategy as suggested in Eqs. 24–25, a different set of $V$, $W$, and $Z$ was used across the echo and training iterations sampled from a random uniform distribution. For training, 200 different sets of $V$, $W$, and $Z$ were precalculated ($B = 200$) before the model training, and one of the sets was randomly selected in each iteration. Python's multi-processing algorithm was used to reduce the computation time for generating the multiple sets and input subspace coefficient images.

*Comparison Methods*

First, to validate the reconstruction performance of the proposed subspace QALAS, it was compared with the conventional QALAS, which used the original five k-space acquisitions for quantitative mapping. The five QALAS images were reconstructed in two different ways for comparisons: 1) parallel imaging with compressed sensing (PI-CS) using $l_1$-wavelet regularization and 2) ZS-SSL, which was modified from the original method (50) to reconstruct five QALAS images jointly using a single deep-learning model. The five QALAS images were concatenated along the channel dimension when fed into the single network. PI-CS with $l_1$-wavelet regularization was implemented with BART (60), and ZS-SSL was implemented using the official code (https://github.com/byaman14/ZS-SSL).

Next, Zero-DeepSub, which is another proposed method in this study, was compared with subspace QALAS using different regularizations: 1) no regularization ('w/o Reg'), 2) LLR regularization ('LLR'), and 3) $l_1$-wavelet-based regularization ('$l_1$-wavelet'). They were implemented with BART (60). The optimal regularization parameters of subspace QALAS with LLR and $l_1$-wavelet were determined by grid searching them in terms of correlation with the reference methods, and image sharpness and denoising. All the reconstruction methods were conducted on 2D multi-slice k-space data, which were constructed by taking the inverse Fourier transform of the acquired 3D k-space data along the frequency encoding direction.

*Signal Dictionary and Low-Rank Subspace Basis*

The signal dictionary was generated based on the QALAS signal model in the *Theory* section with the following $T_1$, $T_2$, $B_1^+$, and IE ranges: $T_1$ = [300–5000 ms], $T_2$ = [10–500 ms], $B_1^+$ = [0.65–1.35], and IE = [0.5–1.0]. A small step size was used for short $T_1$ and $T_2$ values, while it was increased gradually for long $T_1$ and $T_2$ values. Specifically, for $T_1$ values, a 5 ms step size was used for 300–3000 ms and 100 ms for 3000–5000 ms. For $T_2$ values, a 1 ms step size was used for 10–100 ms, 2 ms for 100–200 ms, 10 ms for 200–400 ms, and 20 ms for 400–500 ms. In addition, a 0.05 step size was used for 0.65–1.35 $B_1^+$ values, and a 0.02 step size was used for 0.5–1.0 IE values.

Then, a low-rank subspace basis was calculated by singular value decomposition (SVD), and four bases ($K = 4$) were used, representing simulated signals to within 1.75% error. For *in vivo* experiments, the IE value was set as constant 0.8 for dictionary matching to reduce the number of parameters to be estimated to increase the signal-to-noise ratio (SNR) of the estimated $T_1$, $T_2$, and PD values.

**Image Acquisition**

*Phantom Experiments*

To validate the $T_1$ and $T_2$ accuracy of the conventional and subspace QALAS, phantom experiments were conducted on an ISMRM/NIST system phantom (Serial Number 136) on a 3T MAGNETOM Prisma scanner (Siemens Healthineers, Erlangen, Germany) with a 32ch head receive array. $B_1^+$ maps were acquired using a separate Siemens product turbo-FLASH sequence (61). To match the matrix size with the 3D-QALAS images, the $B_1^+$ maps were interpolated, and the values were thresholded to have the same range as the dictionary. The reference $T_1$ and $T_2$ maps were acquired using IR-FSE and SE-FSE scans, respectively. The detailed scan parameters used for the phantom experiments can be found in Supporting Information Tables S1 and S2. To reduce the computation time and memory requirement for image reconstruction, a software channel compression (SCC) method (62) was used to compress the multi-coil k-space data from 32ch to 16ch.

*In vivo Experiments*

All *in vivo* experiments were conducted with the approval of the Institutional Review Board. *In vivo* data were acquired from four healthy volunteers using a 3D-QALAS sequence on the same 3T scanner with the same 32ch head receive array as in the phantom experiments. Fully sampled 3D-QALAS data was acquired from one subject, whereas prospectively accelerated 3D-QALAS data with an acceleration factor of 2 were acquired from three subjects. All data were retrospectively undersampled for validation. $B_1^+$ maps were acquired using a product turbo-FLASH sequence (61) for $B_1^+$ inhomogeneity correction. To match the matrix size with the 3D-QALAS images, the $B_1^+$ maps were interpolated, and the values were thresholded to have the same range as the dictionary. The detailed scan parameters used for the *in vivo* experiments can be found in Supporting Information Table S1. To reduce the computation time and memory requirement, geometric channel compression (GCC) (63) was used to compress the multi-coil k-space data from 32ch to 8ch.

**Model Comparisons**

The accuracy of the reconstructed $T_1$ and $T_2$ maps using the conventional QALAS and the proposed subspace QALAS with different regularizations was evaluated by comparing them with the reference $T_1$ and $T_2$ maps acquired using IR-FSE and SE-FSE scans using linear regression and Bland–Altman analysis. The $T_1$ values of the eight spheres and $T_2$ values of the seven spheres, which have similar physiological $T_1$ and $T_2$ values of the healthy adult brain tissues ($T_1$ = [600–2500 ms], $T_2$ = [40–350 ms]), on the $T_2$ plate of the ISMRM/NIST system phantom were analyzed by measuring the mean values of the region of interests (ROI) drawn using the ITK-SNAP software (https://www.itksnap.org/) (64).

In order to statistically analyze the phantom results in terms of bias, we hypothesized that subspace QALAS would decrease the $T_1$ and $T_2$ bias compared to conventional QALAS, and made a variable, which calculated the difference between conventional QALAS's percentage difference from the reference method and subspace QALAS's percentage difference from the reference method, for each sphere on the $T_2$ plate of the ISMRM/NIST system phantom. A paired Wilcoxon signed-rank test was performed to test the hypothesis, and the results were considered significant if p-values were less than 0.05.

For *in vivo* analysis, we first analyzed g-factor comparisons between subspace QALAS and conventional QALAS. For g-factor analysis, Monte Carlo simulation (65) was used with 1,000 iterations. To analyze the effectiveness of sampling schemes, uniform and Poisson sampling patterns were compared in terms of the g-factor. Both uniform and Poisson sampling patterns were generated differently across echoes. Uniform sampling was shifted $(\Delta k_y, \Delta k_z) = (0, 0), (1, 0), (0, 1), (1, 1), (1, 0)$ for each echo, respectively. Poisson sampling was generated with slight jittering where reduction factors of $k_y$ and $k_z$ were randomly shifted within 1% in each echo to obtain different sampling patterns.

Then, the reconstructed $T_1$, $T_2$, and PD maps using the conventional QALAS and the proposed subspace QALAS with different regularizations were compared. To evaluate the reconstruction performance with undersampled k-space data, retrospective undersampling was conducted based on the Poisson sampling pattern with different reduction factors. Along with the analysis of the quantitative maps, reconstructed subspace coefficients using different regularizations were compared to analyze the performance difference between the regularizations. A root mean square error (RMSE) metric was used for the evaluation of the reconstructed quantitative maps.

Moreover, synthetic images, including $T_1$w, $T_2$w, $T_2$-fluid-attenuation inversion recovery ($T_2$-FLAIR), magnetization-prepared rapid gradient echo (MPRAGE), and double inversion recovery (DIR) were generated using the reconstructed quantitative maps based on Bloch equations and extended phase graph (EPG) (66). The generated images were compared with the ones acquired using standard sequences. The detailed scan parameters can be found in Supporting Information Table S3.

**RESULTS**

**Phantom Evaluation**

*Accuracy Evaluation of Conventional and Subspace QALAS*

The quantitative $T_1$ and $T_2$ analyses of conventional and subspace QALAS on an ISMRM/NIST system phantom with 3 × 3 reduction factor are shown in Fig. 3. As shown in Fig. 3a–e, both conventional and subspace QALAS show high linearity of $T_1$ values with respect to the reference ones, which were acquired using IR-FSE. The coefficient of determination ($R^2$) of conventional QALAS for $T_1$ values is 0.9865, whereas those of subspace QALAS using without regularization, LLR, $l_1$-wavelet, and Zero-DeepSub are 0.9947, 0.9862, 0.9730, and 0.9857, respectively. In addition, while conventional QALAS shows the regression slope of 1.1369, subspace QALAS using those four different methods, including without regularization, LLR, $l_1$-wavelet, and Zero-DeepSub, show 1.2056, 0.6448, 0.9894, and 1.1596, respectively. Similar results were observed in $T_2$ values, as shown in Fig. 3f–j. The coefficient of determination for $T_2$ values using conventional QALAS is 0.9942, whereas those of subspace QALAS using without regularization, LLR, $l_1$-wavelet, and Zero-DeepSub are 0.9394, 0.9822, 0.9917, and 0.9682, respectively. While the conventional QALAS has a regression slope of 1.3761, subspace using four different regularizations, including without regularization, LLR, $l_1$-wavelet, and Zero-DeepSub, have 1.1589, 1.5474, 1.0486, and 0.8788, respectively.

Fig. 4 shows the Bland–Altman plots of $T_1$ and $T_2$ values acquired using conventional QALAS and subspace QALAS with 3 × 3 reduction factor. The difference values were calculated as percentages. As shown in Fig. 4a–e, compared to conventional QALAS, subspace QALAS with different regularizations, including without regularization, LLR, and $l_1$-wavelet, show higher absolute mean bias from 6.46% up to around 24.66%, where there were significant differences between conventional QALAS and subspace QALAS without regularization and with LLR (Supporting Information Table S4). In contrast, Zero-DeepSub shows slightly reduced limits of agreement and a reduced mean bias of -0.81% compared to conventional QALAS which shows -1.05%, although there was no significant difference between them. The Bland–Altman plots of $T_2$ values are also presented in Fig. 4f–j. Conventional QALAS has a high mean bias of -17.36%, whereas Zero-DeepSub has a significantly reduced mean bias of -1.84% (Supporting Information Table S4) while retaining limits of agreements. Subspace QALAS with $l_1$-wavelet shows -9.70%; however, other subspace QALAS, including without regularization and LLR show higher mean bias of -25.50% and -42.89%, respectively. Compared to the reference method, the $T_2$ values were highly overestimated using conventional QALAS in long $T_2$ values (from 100 to 350 ms), whereas those were mitigated using subspace QALAS.

*Precision Evaluation of Conventional and Subspace QALAS*

Supporting Information Figure S1 shows the coefficient of variation (CoV) (%) analysis of $T_1$ and $T_2$ values using an ISMRM/NIST system phantom with 3 × 3 reduction factor reconstructed using conventional and subspace 3D-QALAS with different regularizations, including without regularization, LLR, $l_1$-wavelet, and Zero-DeepSub. For the $T_1$ analysis, the proposed subspace QALAS with Zero-DeepSub showed the lowest CoV values while other reconstruction methods showed around from 1.9 to 3.7%. For the $T_2$ analysis, subspace QALAS with LLR and Zero-DeepSub had lower CoV values compared to conventional QALAS and other subspace QALAS methods. In all reconstruction methods, the CoV values were between 1.3 and 3.7%.

*Comparison Between Conventional QALAS and Subspace QALAS*

Fig. 5 shows the reconstructed quantitative maps and multi-contrast QALAS images with 3 × 3 reduction factor using conventional and subspace QALAS. $l_1$-wavelet and Zero-DeepSub were used for conventional and subspace QALAS, respectively. As shown in the multi-contrast images, subspace QALAS shows reduced blurring in the reconstructed images, especially in the second one, compared to conventional QALAS. These differences resulted in the reconstructed quantitative $T_1$ and $T_2$ maps. In particular, subspace QALAS showed reduced noise and blurring in the quantitative maps. In addition, in order to see the difference between conventional and subspace QALAS without acceleration, the reconstructed maps and images using fully sampled k-space data are presented in Supporting Information Figure S2.

**In vivo Evaluation**

*Comparison Between Conventional QALAS and Subspace QALAS*

The reconstructed quantitative maps, including $T_1$, $T_2$, and PD maps, and multi-contrast QALAS images with fully sampled k-space data using conventional and subspace QALAS are presented in Supporting Information Figure S3. Here, Zero-DeepSub was used for subspace QALAS reconstruction. Conventional QALAS has blurred tissue contrasts and low contrast-to-noise ratio (CNR) in the reconstructed maps compared to subspace QALAS, as indicated by the blue arrows. The reconstructed multi-contrast images using conventional QALAS are the fully sampled original QALAS images, whereas those reconstructed using subspace QALAS are the first echo of the ETL multi-echo images. As shown in the difference images, there are differences between subspace and conventional QALAS images, especially for the first two contrasts. As shown in the magnified images, the first contrast image of the subspace QALAS shows high CNR compared to conventional QALAS and reduced blurring using subspace QALAS were mostly observed in the second contrast image. These differences resulted in the reconstructed quantitative $T_1$ and $T_2$ maps. In particular, subspace QALAS showed reduced noise in the $T_1$ maps, which made the maps slightly blurrier than the ones of conventional QALAS, whereas T$_2$ maps of subspace QALAS had sharper tissue structures than the ones of conventional QALAS.

*G-factor Analysis*

Fig. 6 shows the g-factor analysis of conventional and subspace QALAS at 3 × 2 reduction factor using (Fig. 6a) uniform sampling and (Fig. 6b) Poisson sampling. For subspace QALAS, the actual reduction factor is $R^* = 24$, which was calculated by multiplying the reduction factor with the number of subspace bases. The g-factor maps of subspace coefficients were calculated for subspace QALAS, whereas individual g-factor maps of each contrast were calculated for conventional QALAS. In both sampling patterns, subspace QALAS presents better g-factor maps of the subspace coefficients than those of the individual contrasts of conventional QALAS. For uniform sampling, subspace QALAS improved $G_{avg}$ and $G_{max}$ by 9.9% and 27.8% on average, respectively, compared to conventional QALAS. For Poisson sampling, subspace QALAS improved $G_{avg}$ and $G_{max}$ by 17.9% and 22.8% on average, respectively, compared to conventional QALAS. The g-factor values demonstrate that Poisson sampling shows better g-factor maps than the uniform sampling and subspace coefficients, and subspace QALAS has better g-factor maps than those of conventional QALAS. It also has been demonstrated that complementary sampling improved g-factor maps in subspace QALAS, whereas the same sampling patterns across echo showed almost identical g-factor maps of the subspace coefficients as the ones of conventional QALAS, as shown in Supporting Information Figure S4.

*3-minute 3D-QALAS for 1mm³ T$_1$ and T$_2$ Mapping*

Fig. 7 shows the reconstructed T$_1$, T$_2$, and PD maps with Poisson sampling and 3 × 2 reduction factor (i.e., 3 min 8 s for 1mm³ isotropic resolution) using the conventional QALAS and subspace QALAS with different regularizations: $l_1$-wavelet and ZS-SSL for conventional QALAS, and without regularization, LLR, $l_1$-wavelet, and Zero-DeepSub for subspace QALAS. Since there are no ground-truth maps for subspace QALAS, difference images and RMSE values were calculated between the maps reconstructed from fully sampled k-space data using each method and those reconstructed from undersampled k-space data using the same method. For reference, the maps reconstructed using subspace QALAS using Zero-DeepSub from fully sampled data were presented. As shown in the difference images and RMSE values, the proposed Zero-DeepSub presented a superior performance in reconstructing $T_1$, $T_2$, and PD maps compared to conventional QALAS and subspace QALAS with other regularizations. In particular, low RMSE values were observed in white matter regions of $T_1$ and PD maps. When the reconstructed maps of conventional QALAS are compared with those of subspace QALAS with the same $l_1$-wavelet regularization, subspace QALAS outperformed conventional QALAS. Fig. 8 shows the ROI analysis of the reconstructed $T_1$ and $T_2$ maps of Fig. 7. Five ROIs drawn on the brain gray matter areas were analyzed, which are presented in Supporting Information Figure S5. Overall, Zero-DeepSub had low standard deviations compared to other methods, especially for $T_2$ values.

To analyze the performance of subspace QALAS, subspace coefficient images are presented in Supporting Information Figure S6 with different regularizations, including without regularization, LLR, $l_1$-wavelet, and Zero-DeepSub. The signal intensities of each subspace coefficient were normalized for visualization. While all subspace QALAS methods demonstrated similar reconstructed first and second, all methods, except for Zero-DeepSub, suffered from noise and residual artifacts in the third and fourth subspace coefficients. Supporting Information Figure S7 shows the reconstructed multi-contrast QALAS images using conventional QALAS along with the sampling masks. The ZS-SSL showed better performance than the $l_1$-wavelet regularization of conventional QALAS; yet, it has residual blurring artifacts, especially in the first QALAS contrast image.

*2-minute 3D-QALAS for 1mm³ $T_1$ and $T_2$ Mapping*

To further push the acceleration, Fig. 9 shows the reconstructed $T_1$, $T_2$, and PD maps with Poisson sampling and 3 × 3 reduction factor (i.e., 2 min 5 s for 1mm³ isotropic resolution) using conventional and subspace QALAS with different regularizations: $l_1$-wavelet and ZS-SSL for conventional QALAS, and without regularization, LLR, $l_1$-wavelet, and Zero-DeepSub for subspace QALAS. Similar to the previous 3-min QALAS experiment, difference images and RMSE values were calculated between the maps reconstructed from fully sampled k-space data using each method and those reconstructed from undersampled k-space data using the same specific method. For reference, the maps reconstructed using subspace QALAS with Zero-DeepSub from fully sampled data were presented. Here, conventional QALAS showed higher biases and errors compared to subspace QALAS. When the reconstructed maps of conventional QALAS are compared with those of subspace QALAS with the same $l_1$-wavelet regularization, subspace QALAS reduced the $T_1$ and PD RMSE values by 2-fold compared to conventional QALAS. Moreover, the ZS-SSL of conventional QALAS showed inferior performance than subspace QALAS with $l_1$-wavelet regularization, as indicated by the $T_1$, $T_2$, and PD RMSE values. In all comparisons, the proposed Zero-DeepSub outperformed other reconstruction methods and low RMSE values were observed in white matter regions of $T_1$ and PD maps. Especially, Zero-DeepSub at 9-fold acceleration was as good as the standard ZS-SSL at 6-fold acceleration. Additional *in vivo* results using Zero-DeepSub are presented in Supporting Information Figure S8.

These results are supported by the reconstruction results of the subspace coefficients and the multi-contrast QALAS images, as shown in Supporting Information Figures S9 and S10. The signal intensities of each subspace coefficient were normalized for visualization. Similar to the 3-min 3D-QALAS experiment, other subspace QALAS methods, except for Zero-DeepSub, suffered from noise and residual artifacts for the third and fourth subspace coefficients. As presented in Supporting Information Figure S10, while the ZS-SSL of conventional QALAS outperformed the one of $l_1$-wavelet regularization, it suffered from remaining artifacts and noise, especially in the first and second contrast QALAS images.

*Synthetic Image Generation Using 3D-QALAS Maps*

Fig. 10 shows the acquired standard $T_1$w, $T_2$w, $T_2$-FLAIR, MPRAGE, and DIR images in the first row and the generated images in the second row, which were generated from the reconstructed quantitative $T_1$, $T_2$, and PD maps with 3-min 3D-QALAS using Zero-DeepSub. While the total scan time required for acquiring the standard images required more than 15 min, subspace QALAS using Zero-DeepSub could generate multi-contrast images with the reconstructed $T_1$, $T_2$, and PD maps from a 3-min 3D-QALAS acquisition. The generated images using 2-min 3D-QALAS using Zero-DeepSub are presented in Supporting Information Figure S11.

**DISCUSSION**

In this work, we first proposed ***"subspace QALAS"*** to reconstruct QALAS time-series images using a low-rank subspace method, thus enabling accurate $T_1$ and $T_2$ mapping with reduced blurring and better g-factor noise mitigation compared to conventional QALAS. Conventional 3D-QALAS assumes that the k-space data are acquired at the first echo of the lengthy ETL using

FLASH readout. The assumption, which neglects $T_1$ relaxation during the five acquisitions within each QALAS block, induces voxel blurring in the reconstructed images and biases in the quantitative $T_1$ and $T_2$ values. Our proposed subspace QALAS, which reconstructs QALAS time-series data using a low-rank subspace method, can utilize full QALAS signal evolution, including $T_1$ relaxation during the readouts, thus enabling accurate estimation of quantitative $T_1$ and $T_2$ values compared to conventional QALAS. To the best of our knowledge, this is the first study to utilize a low-rank subspace method for improving the multiparametric mapping using a 3D-QALAS sequence.

Moreover, we proposed a novel zero-shot deep subspace method called ***"Zero-DeepSub"*** to further improve the fidelity of subspace QALAS by combining a scan-specific deep-learning-based reconstruction and low-rank subspace modeling. Zero-DeepSub reconstructs denoised subspace coefficients from the acquired multi-echo k-space that can be used for generating quantitative maps. While many deep-learning-based studies have been focused on reconstructing undersampled MR images or qMRI using a supervised learning scheme, there are several cases where fully sampled or reference quantitative maps are hard to be obtained or defined. In particular, our proposed subspace QALAS is one of those cases where fully sampled k-space acquisition for over 600 echoes (e.g., ETL × 5 > 600) is not feasible using 3D-FLASH readouts. A scan-specific or unsupervised learning scheme needs to be used for subspace QALAS while leveraging deep-learning-based regularization, which has been known to have superior performance than conventional hand-crafted regularizations such as wavelet or total variation.

The ISMRM/NIST system phantom results demonstrated that both conventional and subspace QALAS showed good linearity with respect to the reference methods, including IR-FSE and SE-FSE, in terms of $T_1$ and $T_2$ values. In particular, subspace QALAS with Zero-DeepSub could reduce the $T_2$ mean bias down to 1.84%, whereas conventional QALAS had over 17% mean bias, as shown in Fig. 4. The key assumption of conventional QALAS that there was no $T_1$ relaxation during the lengthy FLASH readouts might cause inaccurate $T_1$ and $T_2$ estimation using the truncated five-point dictionary. The proposed subspace QALAS could fully utilize QALAS signal evolution, including signal relaxation during lengthy readouts, and could reduce the biases of the estimated quantified values. In addition, considering the CoV analysis presented in Supporting Information Figure S1, Zero-DeepSub improved accuracy while showing improved precision compared to that of conventional QALAS.

The purpose of using Bland–Altman plot analysis is to assess bias among mean variances and to estimate a confidence interval of agreement, encompassing 95% of the variances in the second method when compared to the first (67,68). As such, Bland–Altman analysis did not involve a hypothesis test or p-value (67,68). Instead, in order to statistically analyze the quantitative results of the phantom results in terms of bias, a paired Wilcoxon signed-rank test was performed to test whether subspace QALAS would decrease the $T_1$ and $T_2$ bias compared to conventional QALAS. As shown in Figure 4 and Supporting Information Table S4, the results demonstrated that there was a significant difference between the $T_2$ bias of conventional QALAS and Zero-DeepSub, where Zero-DeepSub had a reduced mean bias of -1.84% whereas conventional QALAS had -17.36%.

According to the *in vivo* results, the proposed subspace QALAS presented more robust performance than conventional QALAS with high reduction factors. In particular, using the same $l_1$-wavelet regularization, subspace QALAS showed lower RMSE errors than the ones of conventional QALAS, demonstrating the effectiveness of subspace QALAS utilizing the full QALAS signal evolution. Another proposed method, Zero-DeepSub, outperformed other regularizations, including $l_1$-wavelet and scan-specific deep learning method (ZS-SSL). In particular, Zero-DeepSub showed better subspace coefficients while other subspace regularizations suffered from residual noise and aliasing artifacts, specifically for the third and fourth coefficients. Thus, Zero-DeepSub could further push the reduction factor up to 9-fold, enabling whole-brain $T_1$, $T_2$, and PD mapping at 1 mm isotropic resolution within 2 min. Zero-DeepSub with 9-fold acceleration was as good as standard ZS-SSL with 6-fold acceleration: Zero-DeepSub with 9-fold acceleration had (12.00 and 14.99% for $T_1$, 12.30 and 12.17% for $T_2$, and 5.46 and 8.62% for PD) white matter and gray matter RMSE values whereas standard ZS-SSL with 6-fold acceleration had (13.31 and 18.14% for $T_1$, 10.69 and 11.10% for $T_2$, and 6.53 and 11.03% for PD). The reconstructed quantitative maps and subspace coefficients from fully-sampled data are presented in Supporting Information Figures S12 and S13.

For g-factor analysis, the g-factor maps of subspace coefficients were calculated for subspace QALAS, whereas individual g-factor maps of each contrast were calculated for conventional QALAS. While conventional QALAS reconstructed each contrast image individually, subspace QALAS, which utilized time-resolved reconstruction with subspace basis, reconstructed subspace coefficient images from the acquired k-space data, as defined in Eqs. 17–18; thus, the g-factor for subspace QALAS needed to be defined on the subspace coefficients. The g-factor analysis showed that Poisson sampling was beneficial for both conventional and subspace QALAS, and subspace QALAS had better g-factor maps of the subspace coefficients than conventional QALAS. While the individual QALAS contrast image had similar g-factor maps, the subspace coefficients utilizing time-resolved reconstruction with subspace basis had reduced g-factor values, and the first coefficient, which was the dominant coefficient for time-resolved reconstruction particularly, showed the lowest g-factor values on average. Moreover, while there was no improvement if the same sampling pattern was used across echoes, subspace QALAS had g-factor improvements using complementary sampling compared to conventional QALAS. These analyses supported that subspace QALAS was more robust than conventional QALAS with high reduction factors. They also demonstrated that subspace QALAS had improved g-factor maps due to better reconstruction conditions even though the decomposed k-space data at each time point was highly undersampled (>100 reduction factor) in terms of the conventional subspace or shuffling methods.

The computation time of Zero-DeepSub was reduced by changing the matrix multiplication order and using Python's multi-processing algorithm. First, the conjugate gradient algorithm-based data consistency layer in Zero-DeepSub requires high computation since it requires the repeated computation of forward and adjoint operations of **A**. For subspace QALAS, it at least requires $T$ (the number of echoes) $\times$ $L$ (the number of coils) $\approx 5000$ Fourier transform along with $T \times K$ (the number of subspace basis) $\approx 2500$ matrix multiplication computation for each voxel. However, by using a similar approach of the shuffling method (33), the computation of Fourier transform can be reduced from $T \times L \approx 5000$ to $K \times L \approx 30$, and the computation of matrix multiplication can also be reduced from $T \times K \approx 2500$ to $K \times K = 16$ element-wise multiplication. Second, multiple sets of the decomposed k-space data and input subspace coefficient images needed to be pre-calculated before model training. The computation time was reduced by using Python's multi-processing algorithm, which parallelizes the computation with multiple Python processes. By utilizing those tricks, the model training could be done on a single GPU and took about 4 h for a multi-slice whole-brain reconstruction. To accelerate the computation further, parallel GPU computing with multiple GPUs and deep-learning training schemes, including pre-training and transfer learning, can be utilized. For instance, a self-supervised-learning-based quantitative mapping method, which was proposed for rapid mapping of multiparametric maps instead of a dictionary matching method, could reduce the parameter estimation time for QALAS from 1.5 h scan-specific training to 15 min fine-tuning using a transfer learning scheme (69). Similar computation reduction can be expected for Zero-DeepSub utilizing those training schemes.

When an adiabatic inversion RF pulse is used, the inversion of longitudinal magnetization can be incomplete because of $T_2$ relaxation during a lengthy RF pulse, which can be quantified as IE (70). The 3D-QALAS sequence uses an adiabatic inversion RF pulse before the second FLASH readout; thus, IE needs to be quantified by including it in both the subspace basis for reconstructing time-series images and the signal dictionary for mapping. Recent studies conducted phantom experiments on an ISMRM/NIST system phantom using a 3D-QALAS sequence and demonstrated that an IE estimation along with $T_1$, $T_2$, and PD values could increase the accuracy of the $T_1$ and $T_2$ values when comparing to the reference values (30,69). While IE estimation could help to increase the accuracy of the $T_1$ and $T_2$ values, the increased number of parameters to be estimated by including the IE can increase the noise sensitivity and decrease the SNR of the estimated maps. In our experiments, IE values were included in the subspace basis and dictionary matching for the phantom experiments. For the *in vivo* experiments, while they were included in the subspace basis, they were set as a constant 0.8 value in the dictionary matching to reduce the degree of freedom of the dictionary matching and increase the SNR of the estimated $T_1$, $T_2$, and PD maps. The value was guided by the mean of the estimated *in vivo* IE maps when they were estimated for the full 0.5–1.0 values using dictionary matching. Another way to estimate the IE values is to use a Bloch simulation for the adiabatic inversion pulse for any given $T_1$ and $T_2$ pair. Moreover, RF pulse shapes and pulse duration can further be optimized using Bloch simulation, which we plan to pursue in future work. Since 3D-QALAS uses non-slice-selective RF pulses, we did not have to consider slice profile selection in this work.

The *in vivo* results demonstrated that multi-contrast synthetic images could be generated from the quantitative $T_1$, $T_2$, and PD maps acquired using subspace QALAS; yet, the $T_2$-FLAIR image needs more improvements in terms of contrast. Other synthetic MRI studies also have reported similar issues, including inferior contrast or SNR, hyperintense signals around the brain surface, and partial volume effect, for $T_2$-FLAIR (7,71–73). Partial volume artifacts can be reduced by using multi-component analysis (74,75), which utilizes that the signal of a single voxel can be represented by the weighted summation of multiple tissue components. In future works, we will adapt the multi-component analysis for 3D-QALAS, which can improve the quality of the generated contrast images with reduced blurring and partial volume artifacts.

A recent study using 3D-MRF using multi-axis spiral projection could enable whole-brain $T_1$, $T_2$, and PD mapping at 1 mm isotropic resolution within 2 min (76). MRF or MR-STAT encodes the tissue parameters using randomized flip angle trains or repetition times, which aims to create a unique signal evolution from different tissues (18,77). On the other hand, 3D-QALAS has an adiabatic $T_2$ preparation pulse and an inversion pulse, respectively, encoding the $T_1$- and $T_2$-sensitized signals. Supporting Information Figure S14 presents signal evolution and subspace bases of QALAS and MRF. MRF signal evolutions were generated based on the MRF toolbox given by (76) for a FISP acquisition using the same $T_1$ and $T_2$ ranges used for QALAS. In QALAS, four bases represent simulated signals to within 1.75% RMSE, whereas in MRF, four bases showed 4.12% RMSE, and five bases yielded 2.44% RMSE. This implies that MRF requires a greater number of subspace bases for representing the original signals compared to QALAS. In addition, MRF utilizes rapid spiral readouts, which might be vulnerable to gradient system imperfections, including eddy currents (77). In contrast, our proposed method based on a 3D-QALAS sequence uses Cartesian readouts that do not require nonuniform Fourier transforms, $B_0$, or trajectory correction. Non-cartesian readouts are inherently efficient, but QALAS also lends itself to wave-encoding that similarly boosts efficiency, which can be further accelerated with the Zero-DeepSub scheme to achieve 2 min acquisitions (29).

Although there was no motion artifact observed in our acquired data, 3D-QALAS with fully sampled acquisitions might be vulnerable to motion artifacts due to the long scan time. While the proposed subspace QALAS was designed to reduce the scan time, another way to mitigate motion is to apply prospective motion correction with a spiral navigator, as in (78).

Our proposed Zero-DeepSub method can be applied to other qMRI sequences, such as 3D-MRF or EPTI, which usually uses LLR or wavelet-based regularizations, and can improve the fidelity of the quantitative maps by regularizing them with the scan-specific deep subspace model. Moreover, a clinical study using subspace QALAS can be conducted to translate our proposed methods.

**CONCLUSION**

We proposed a subspace QALAS to reconstruct QALAS time-series images using a low-rank subspace method, thus enabling accurate $T_1$ and $T_2$ mapping with reduced blurring compared to conventional QALAS. We also proposed Zero-DeepSub, a novel zero-shot deep subspace method, to further improve the fidelity of subspace QALAS by combining a scan-specific deep-learning-based reconstruction method and a low-rank subspace method. Our subspace QALAS with Zero-DeepSub improved RMSE values up to 2-fold compared to standard subspace regularizers and enabled whole-brain $T_1$, $T_2$, and PD mapping at 1 mm isotropic resolution within 2 min of scan time. This technique has great potential to substantially reduce scan times associated with clinical MRI exams while providing quantitative tissue characterization.

**ACKNOWLEDGMENTS**

This work was supported by research grants NIH R01 EB032378, R01 EB028797, R03 EB031175, P41 EB030006, U01 EB026996, U01 DA055353, UG3 EB034875 and the NVIDIA Corporation for computing support.

**DATA AVAILABILITY STATEMENT**

All the source codes can be found here: https://github.com/yohan-jun/Zero-DeepSub

**Figure Legends**

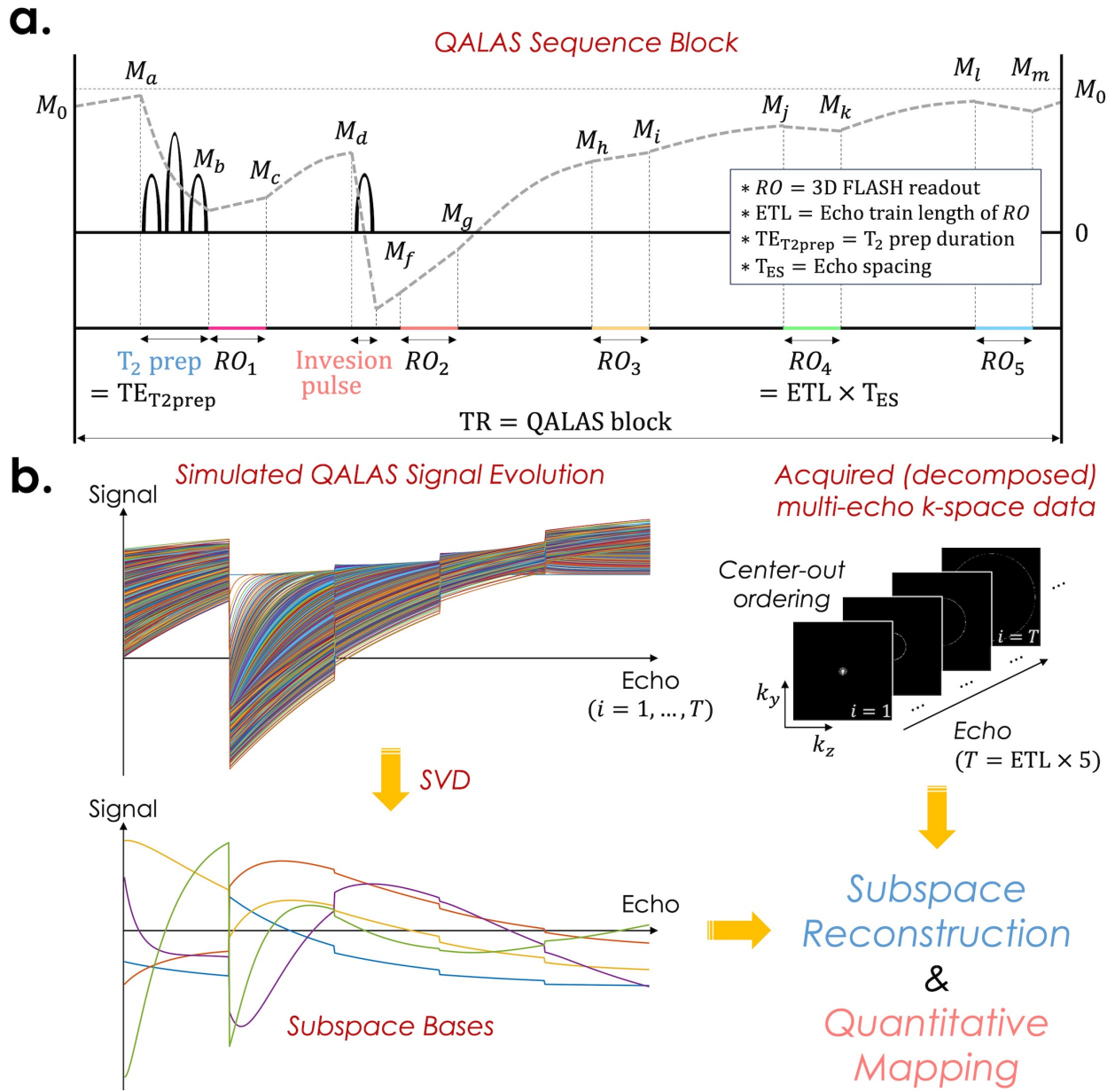

Figure 1. (a) Sequence diagram of 3D-quantification using an interleaved Look-Locker acquisition sequence with $T_2$ preparation pulse (3D-QALAS), which has a $T_2$ preparation pulse before the first acquisition and an inversion pulse before the second acquisition. (b) Overall reconstruction scheme of the proposed subspace QALAS. A low-rank subspace reconstruction can be done using the subspace bases, which are calculated from the simulated QALAS signal evolution using a singular

value decomposition method, and quantitative maps, including $T_1$, $T_2$, and proton density maps, can be acquired using a dictionary matching method from the reconstructed time-series images.

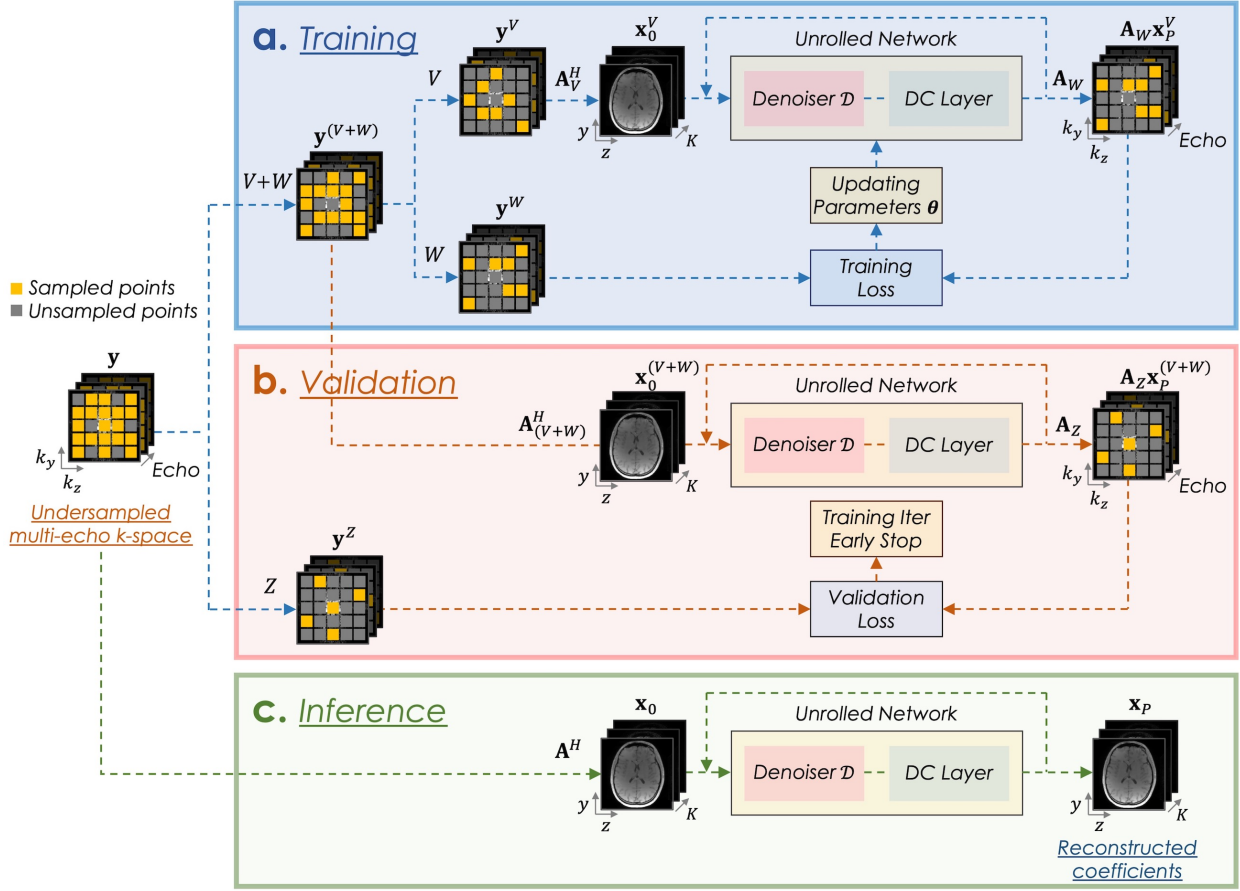

Figure 2. Overall framework of the proposed zero-shot deep-learning subspace method (i.e., Zero-DeepSub). The acquired undersampled multi-echo k-space data can be divided into three subsets using sampling locations (V, W, and Z) for model training and validation. (a) For training, sampling location V is used for data consistency in the unrolled network, and the other sampling location W is used for training loss calculation to update the model parameters. (b) For validation, the combined sampling location V+W is used for data consistency in the unrolled network, and the remaining sampling location Z is used for validation loss calculation to early stop the model training. (c) For model inference, the original acquired k-space data is used for final reconstruction of the subspace coefficients.

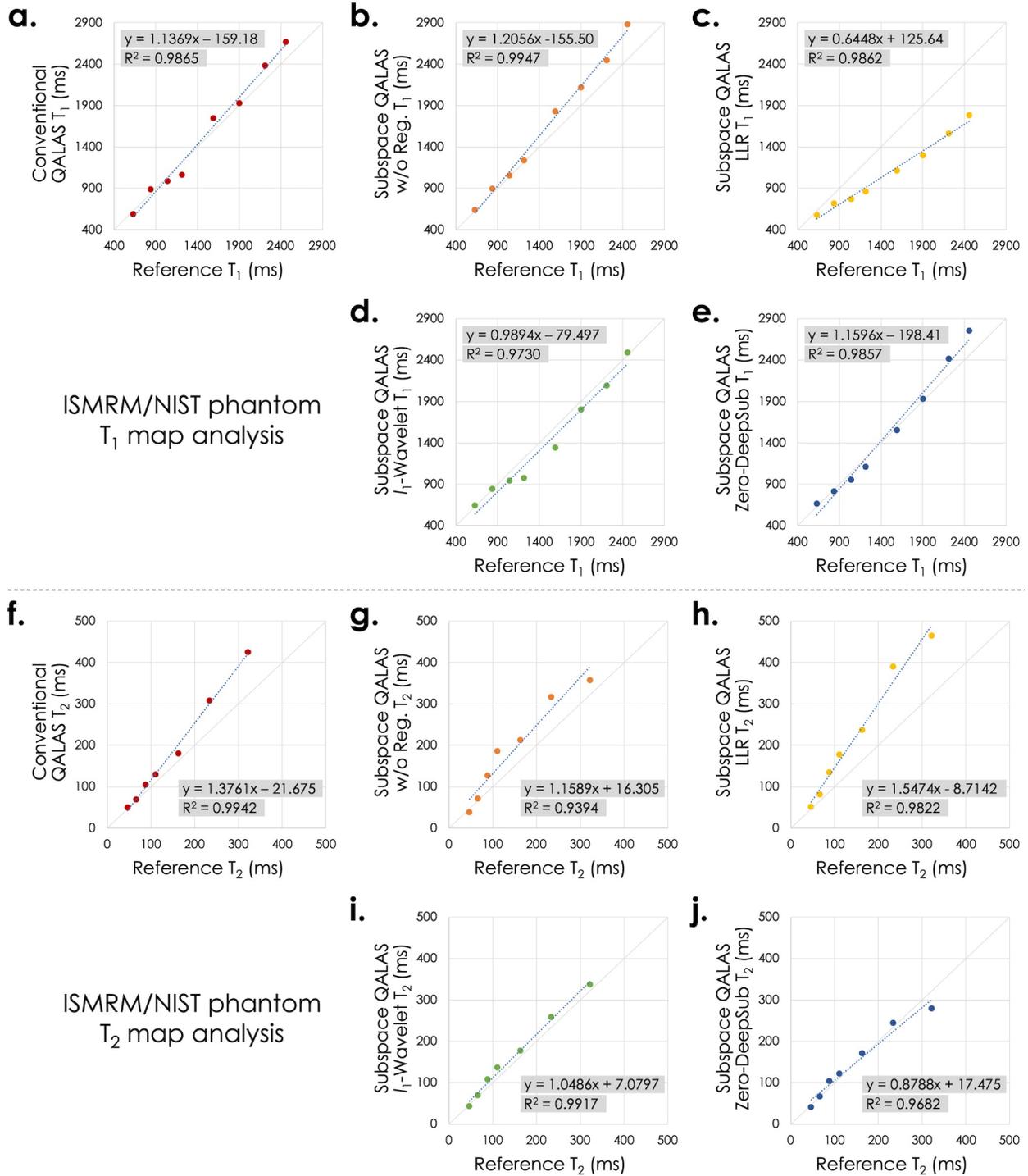

Figure 3. Quantitative $T_1$ and $T_2$ analyses of conventional and subspace 3D-QALAS methods on an ISMRM/NIST system phantom with 3 × 3 reduction factor. Comparisons of $T_1$ and $T_2$ values between the reference methods (IR-FSE and SE-FSE) and (a, f) conventional QALAS, (b, g) subspace QALAS without regularization, (c, h) subspace QALAS with locally low-rank (LLR) regularization, (d, i) subspace QALAS with $l_1$-wavelet regularization, and (e, j) subspace QALAS

with Zero-DeepSub. 3D-QALAS: 3D-quantification using an interleaved Look-Locker acquisition sequence with $T_2$ preparation pulse; ISMRM/NIST: International Society for Magnetic Resonance in Medicine and National Institute of Standards and Technology; IR-FSE: inversion-recovery fast-spin-echo; SE-FSE: single-echo fast-spin-echo.

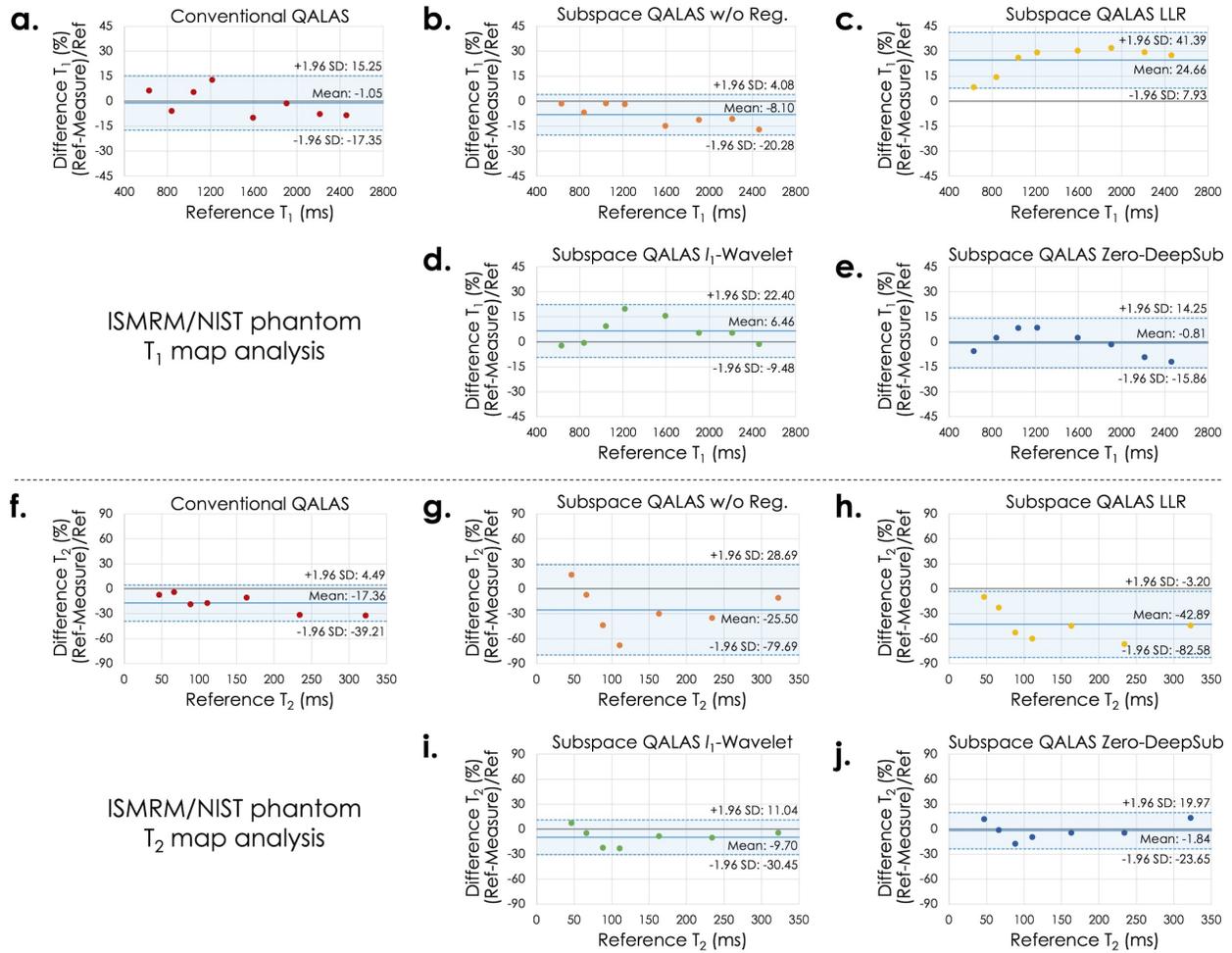

Figure 4. Bland–Altman plots of $T_1$ and $T_2$ values using an ISMRM/NIST system phantom with 3 × 3 reduction factor comparing between the reference methods (IR-FSE and SE-FSE) and (a, f) conventional 3D-QALAS, (b, g) subspace QALAS without regularization, (c, h) subspace QALAS with locally low-rank (LLR) regularization, (d, i) subspace QALAS with $l_1$-wavelet regularization, and (e, j) subspace QALAS with Zero-DeepSub. The difference values were calculated as percentages. Dotted lines indicate 95% limits of agreement calculated as the 1.96 standard deviation of the difference between the reference $T_1$ and $T_2$ values and the ones of the comparison method. 3D-QALAS: 3D-quantification using an interleaved Look-Locker acquisition sequence with $T_2$ preparation pulse; ISMRM/NIST: International Society for Magnetic Resonance in Medicine and National Institute of Standards and Technology; IR-FSE: inversion-recovery fast-spin-echo; SE-FSE: single-echo fast-spin-echo.

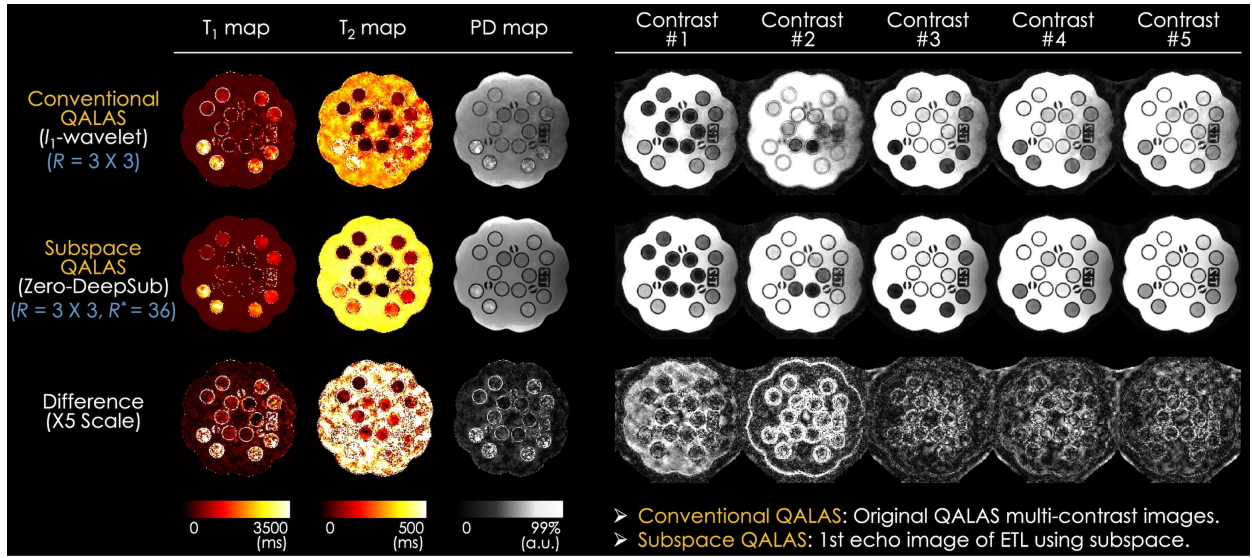

Figure 5. Reconstructed quantitative maps, including $T_1$, $T_2$, and proton density (PD) maps, and multi-contrast 3D-QALAS images with 3 × 3 reduction factor using conventional and subspace 3D-QALAS with an ISMRM/NIST system phantom. For subspace QALAS, the actual reduction factor is $R^* = 36$, which was calculated by multiplying the reduction factor with the number of subspace basis. $l_1$-wavelet and Zero-DeepSub were used for conventional and subspace QALAS, respectively. The reconstructed multi-contrast images using conventional QALAS are original QALAS images, whereas those reconstructed using subspace QALAS are the first echo of the echo train length (ETL) multi-echo images. 3D-QALAS: 3D-quantification using an interleaved Look-Locker acquisition sequence with $T_2$ preparation pulse; ISMRM/NIST: International Society for Magnetic Resonance in Medicine and National Institute of Standards and Technology.

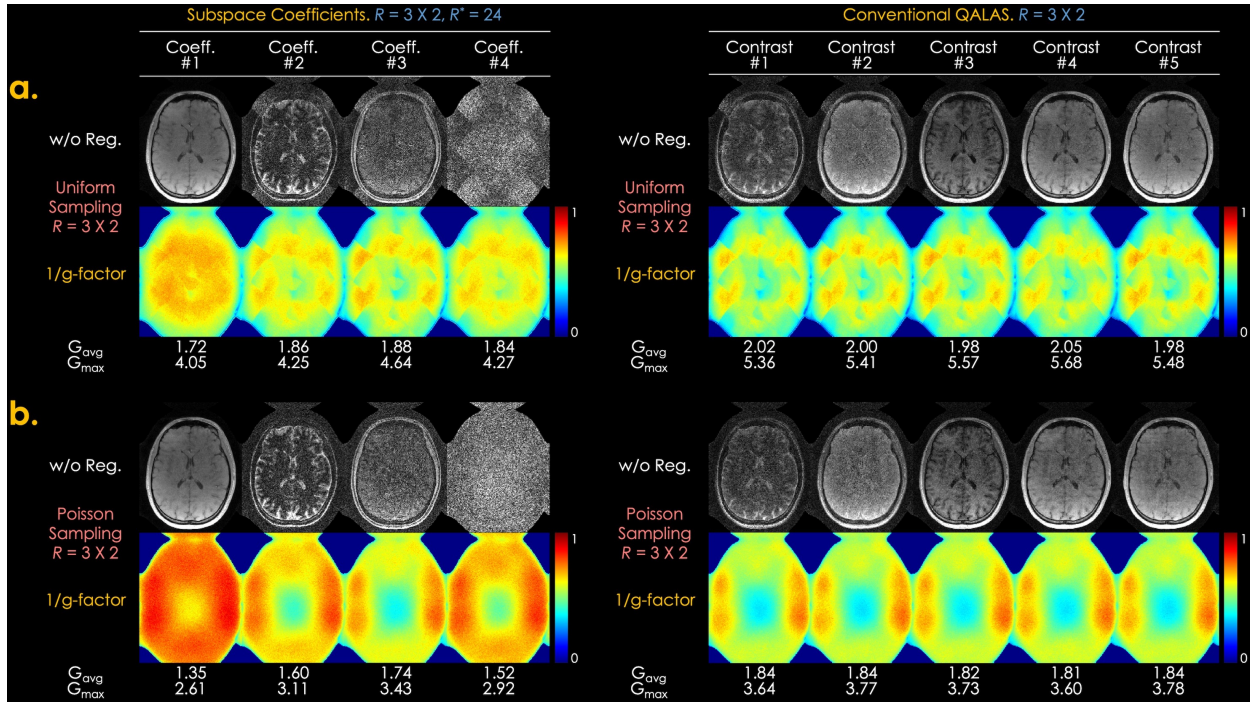

Figure 6. G-factor analysis of conventional and subspace 3D-QALAS with 3 × 2 reduction factor using (a) uniform sampling and (b) Poisson sampling patterns. For subspace QALAS, the actual reduction factor is $R^* = 24$, which was calculated by multiplying the reduction factor with the number of subspace basis. The g-factor maps of subspace coefficients were calculated for subspace QALAS, whereas individual g-factor maps of each contrast image were calculated for conventional QALAS. 3D-QALAS: 3D-quantification using an interleaved Look-Locker acquisition sequence with $T_2$ preparation pulse.

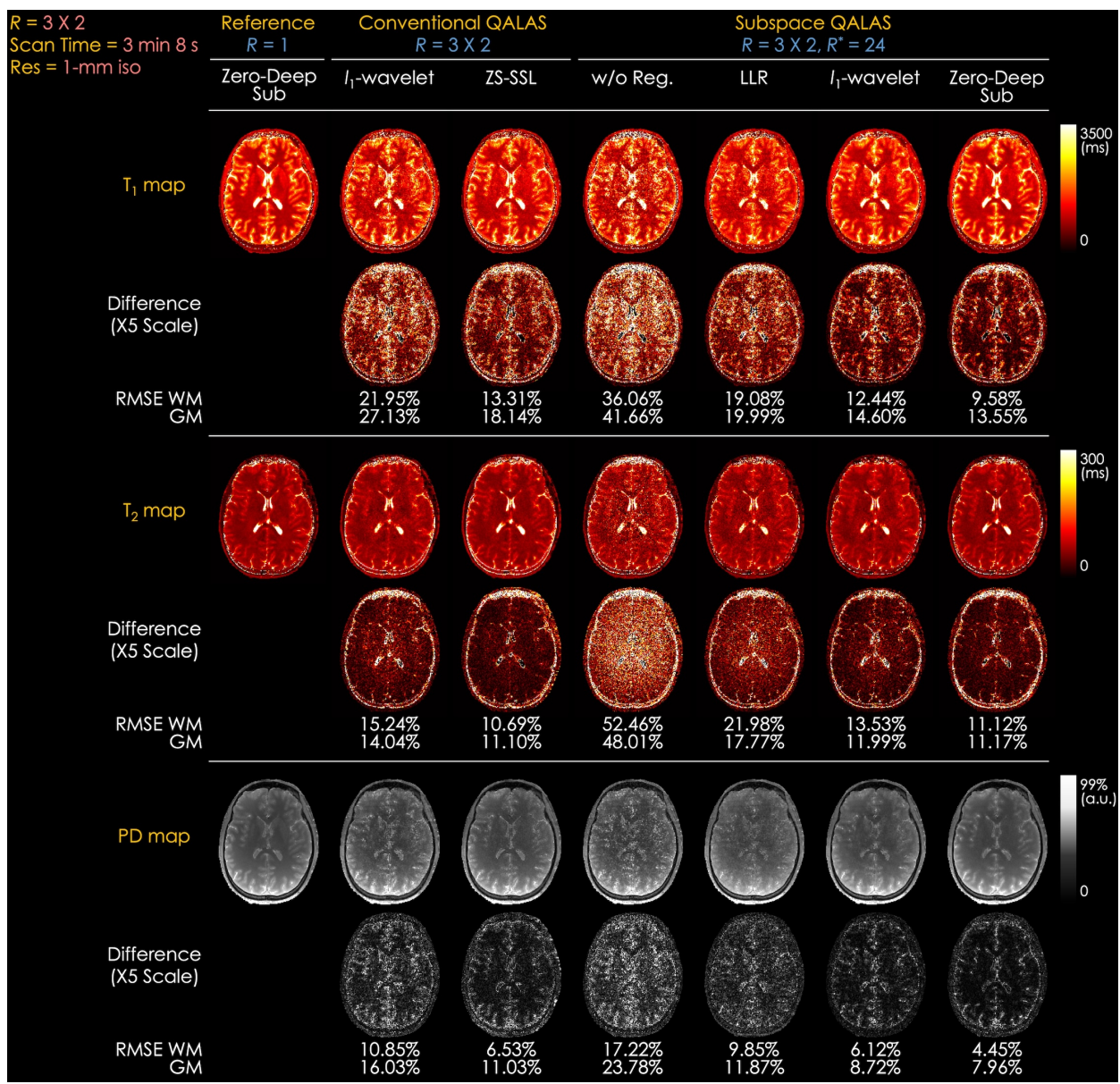

Figure 7. Reconstructed $T_1$, $T_2$, and proton density (PD) maps with $3 \times 2$ reduction factor using Poisson sampling, which enables 3 min 8 s scan time for 1mm$^3$ isotropic resolution, using conventional and subspace 3D-QALAS with different regularizations: $l_1$-wavelet and ZS-SSL for conventional QALAS, and without regularization, LLR, $l_1$-wavelet, and Zero-DeepSub for subspace QALAS. For subspace QALAS, the actual reduction factor is $R^* = 24$, which was calculated by multiplying the reduction factor with the number of subspace basis. Difference images and RMSE values were calculated between the maps reconstructed from fully sampled k-space data using each method and those reconstructed from undersampled k-space data. 3D-

QALAS: 3D-quantification using an interleaved Look-Locker acquisition sequence with $T_2$ preparation pulse; RMSE: root mean square error; WM: white matter; GM: gray matter.

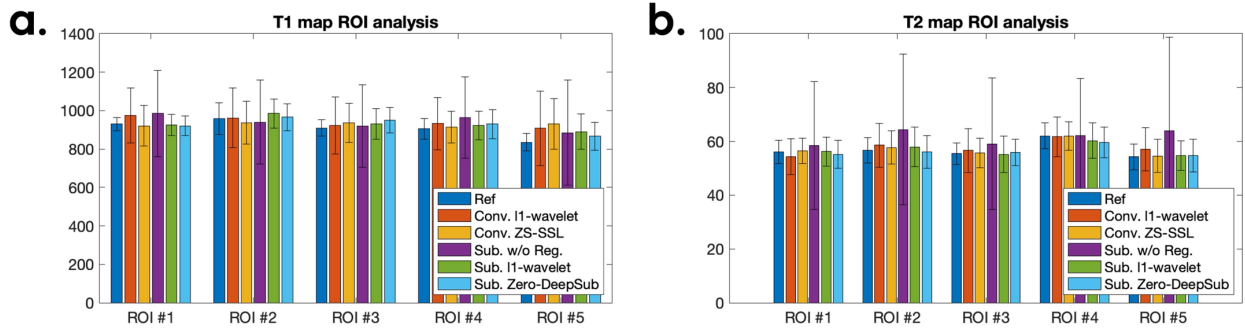

Figure 8. Region of interest (ROI) analysis of the reconstructed *in vivo* (a) $T_1$ and (b) $T_2$ maps with $3 \times 2$ reduction factor using conventional and subspace 3D-QALAS with different regularizations: $l_1$-wavelet and ZS-SSL for conventional QALAS, and without regularization, $l_1$-wavelet, and Zero-DeepSub for subspace QALAS. Five ROIs drawn on the brain gray matter areas are analyzed. 3D-QALAS: 3D-quantification using an interleaved Look-Locker acquisition sequence with $T_2$ preparation pulse.

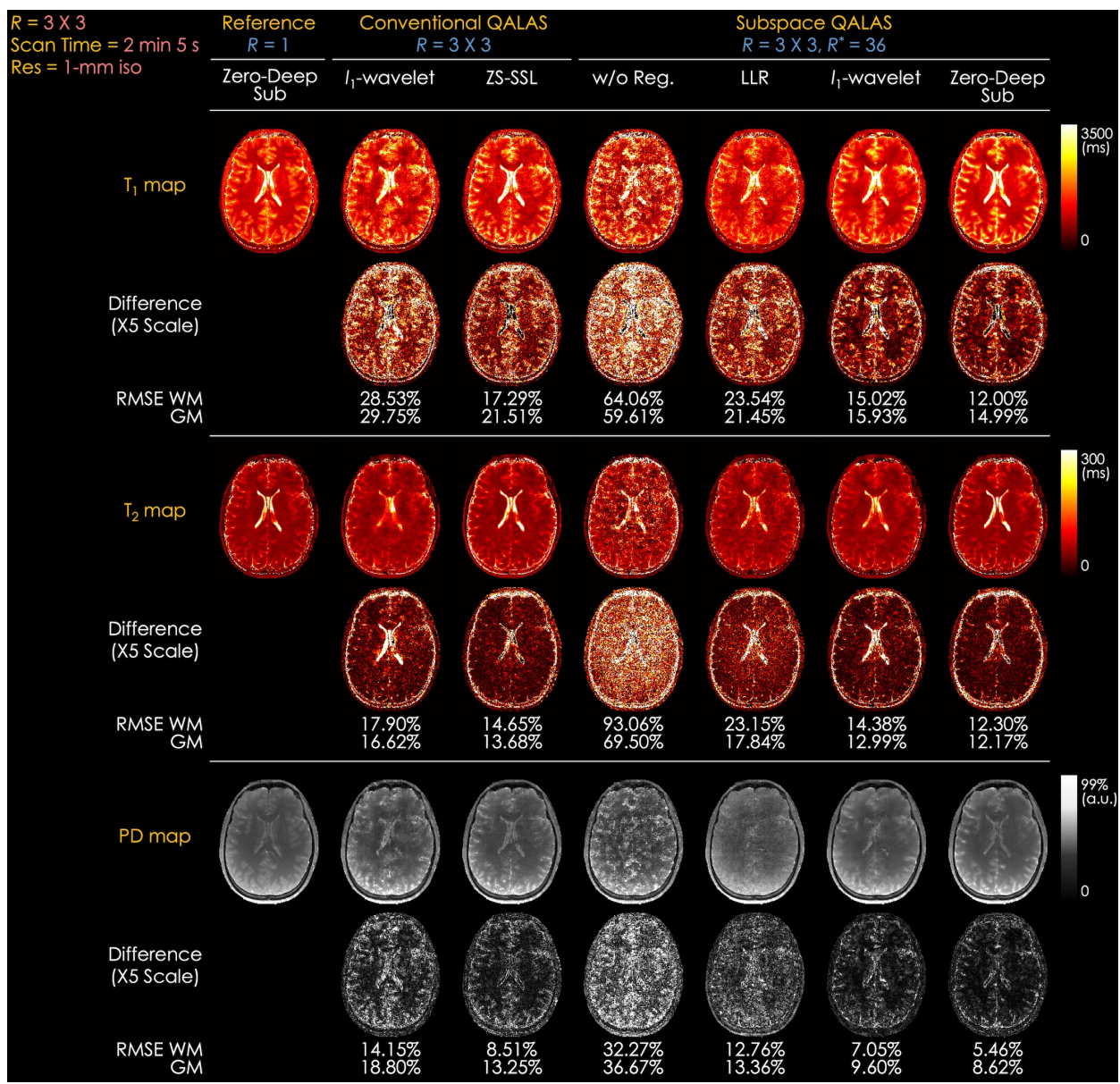

Figure 9. Reconstructed $T_1$, $T_2$, and proton density (PD) maps with $3 \times 3$ reduction factor using Poisson sampling, which enables 2 min 5 s scan time for 1mm³ isotropic resolution, using conventional and subspace 3D-QALAS with different regularizations: $l_1$-wavelet and ZS-SSL for conventional QALAS, and without regularization, LLR, $l_1$-wavelet, and Zero-DeepSub for subspace QALAS. For subspace QALAS, the actual reduction factor is $R^* = 36$, which was calculated by multiplying the reduction factor with the number of subspace basis. Difference images and RMSE values were calculated between the maps reconstructed from fully sampled k-space data using each method and those reconstructed from undersampled k-space data. 3D-

QALAS: 3D-quantification using an interleaved Look-Locker acquisition sequence with $T_2$ preparation pulse; RMSE: root mean square error; WM: white matter; GM: gray matter.

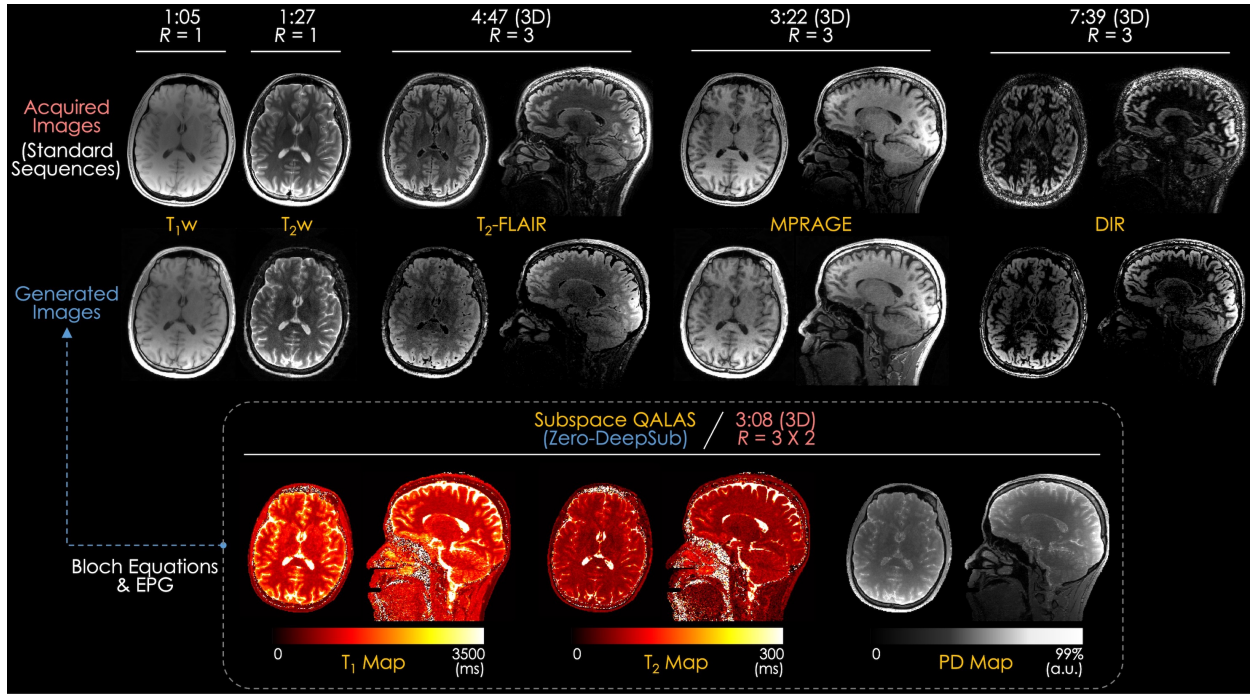

Figure 10. Acquired and generated $T_1$w, $T_2$w, $T_2$-FLAIR, MPRAGE, and DIR images. The images were generated from the reconstructed quantitative $T_1$, $T_2$, and proton density (PD) maps with 3-min 3D-QALAS using Zero-DeepSub based on Bloch equations and extended phase graph (EPG). $T_2$-FLAIR: $T_2$-fluid-attenuation inversion recovery; MPRAGE: magnetization-prepared rapid gradient echo; DIR: double inversion recovery; 3D-QALAS: 3D-quantification using an interleaved Look-Locker acquisition sequence with $T_2$ preparation pulse.

**Supporting Information Legends**

Supporting Information Table S1. MRI scan parameters of 3D-QALAS and turbo-FLASH sequences used for ISMRM/NIST system phantom and *in vivo* experiments.

Supporting Information Table S2. MRI scan parameters of IR-FSE and SE-FSE sequences used for ISMRM/NIST system phantom experiment.

Supporting Information Table S3. MRI scan parameters of $T_1$w, $T_2$w, $T_2$-FLAIR, MPRAGE, and DIR sequence used for *in vivo* experiments.

Supporting Information Table S4. P-values calculated from $T_1$ and $T_2$ bias comparison between conventional 3D-QALAS and subspace QALAS used in ISMRM/NIST system phantom experiment.

Supporting Information Figure S1. Coefficient of variation (CoV) (%) analysis of (a) $T_1$ and (b) $T_2$ values using an ISMRM/NIST system phantom with 3 × 3 reduction factor reconstructed using conventional and subspace 3D-QALAS with different regularizations, including without regularization, locally low-rank (LLR), $l_1$-wavelet, Zero-DeepSub without phase normalization, and Zero-DeepSub with phase normalization. ISMRM/NIST: International Society for Magnetic Resonance in Medicine and National Institute of Standards and Technology; 3D-QALAS: 3D-quantification using an interleaved Look-Locker acquisition sequence with $T_2$ preparation pulse.

Supporting Information Figure S2. Reconstructed quantitative maps, including $T_1$, $T_2$, and proton density (PD) maps, and multi-contrast 3D-QALAS images using conventional and subspace 3D-QALAS with an ISMRM/NIST system phantom. For subspace QALAS, the actual reduction factor is $R^* = 4$, which was calculated by multiplying the reduction factor with the number of subspace basis. $l_1$-wavelet and Zero-DeepSub were used for conventional and subspace QALAS, respectively. The reconstructed multi-contrast images using conventional QALAS are original QALAS images, whereas those reconstructed using subspace QALAS are the first echo of the echo train length (ETL) multi-echo images. 3D-QALAS: 3D-quantification using an interleaved Look-Locker acquisition sequence with $T_2$ preparation pulse; ISMRM/NIST: International Society for Magnetic Resonance in Medicine and National Institute of Standards and Technology.

Supporting Information Figure S3. Reconstructed quantitative maps, including $T_1$, $T_2$, and proton density (PD) maps, and multi-contrast 3D-QALAS images using conventional and subspace 3D-QALAS. For subspace QALAS, the actual reduction factor is $R^* = 4$, which was calculated by multiplying the reduction factor with the number of subspace basis. $l_1$-wavelet and Zero-DeepSub were used for conventional and subspace QALAS, respectively. The reconstructed multi-contrast images using conventional QALAS are the fully sampled original QALAS images, whereas those reconstructed using subspace QALAS are the first echo of the echo train length (ETL) multi-echo images. 3D-QALAS: 3D-quantification using an interleaved Look-Locker acquisition sequence with $T_2$ preparation pulse.

Supporting Information Figure S4. G-factor analysis of conventional and subspace 3D-QALAS with 3 × 2 reduction factor using (a) uniform without elliptical and echo-shift sampling and (b) uniform without elliptical but with echo-shift sampling patterns. For subspace QALAS, the actual reduction factor is $R^* = 24$, which was calculated by multiplying the reduction factor with the number of subspace basis. The g-factor maps of subspace coefficients were calculated for subspace QALAS, whereas individual g-factor maps of each contrast image were calculated for conventional QALAS. 3D-QALAS: 3D-quantification using an interleaved Look-Locker acquisition sequence with $T_2$ preparation pulse.

Supporting Information Figure S5. Region of interests (ROIs) used for quantitative analysis of $T_1$ and $T_2$ maps for 3D-QALAS. 3D-QALAS: 3D-quantification using an interleaved Look-Locker acquisition sequence with $T_2$ preparation pulse.

Supporting Information Figure S6. Reconstructed subspace coefficient images with Poisson sampling and 3 × 2 reduction factor using subspace 3D-QALAS with different regularizations, including without regularization, LLR, $l_1$-wavelet, and Zero-DeepSub. For subspace QALAS, the actual reduction factor is $R^* = 24$, which was calculated by multiplying the reduction factor with the number of subspace basis. The signal intensities of each subspace coefficient were normalized for visualization. 3D-QALAS: 3D-quantification using an interleaved Look-Locker acquisition sequence with $T_2$ preparation pulse.

Supporting Information Figure S7. Reconstructed multi-contrast QALAS images with Poisson sampling and 3 × 2 reduction factor using conventional 3D-QALAS with different regularizations, including $l_1$-wavelet and ZS-SSL, along with the sampling masks. 3D-QALAS: 3D-quantification using an interleaved Look-Locker acquisition sequence with $T_2$ preparation pulse.

Supporting Information Figure S8. Reconstructed $T_1$, $T_2$, and proton density (PD) maps with 3 × 3 reduction factor using Poisson sampling, which enables 2 min 5 s scan time for $1mm^3$ isotropic resolution, using subspace 3D-QALAS with Zero-DeepSub. The data were acquired with acceleration factor 2, and additional retrospective undersampling was conducted based on the Poisson sampling pattern to get 3 × 3 reduction factor. 3D-QALAS: 3D-quantification using an interleaved Look-Locker acquisition sequence with $T_2$ preparation pulse.

Supporting Information Figure S9. Reconstructed subspace coefficient images with Poisson sampling and 3 × 3 reduction factor using subspace 3D-QALAS with different regularizations, including without regularization, LLR, $l_1$-wavelet, and Zero-DeepSub. For subspace QALAS, the actual reduction factor is $R^* = 36$, which was calculated by multiplying the reduction factor with the number of subspace basis. The signal intensities of each subspace coefficient were normalized for visualization. 3D-QALAS: 3D-quantification using an interleaved Look-Locker acquisition sequence with $T_2$ preparation pulse.

Supporting Information Figure S10. Reconstructed multi-contrast QALAS images with Poisson sampling and 3 × 3 reduction factor using conventional 3D-QALAS with different regularizations, including $l_1$-wavelet and ZS-SSL, along with the sampling masks. 3D-QALAS: 3D-quantification using an interleaved Look-Locker acquisition sequence with $T_2$ preparation pulse.

Supporting Information Figure S11. Acquired and generated $T_1$w, $T_2$w, $T_2$-FLAIR, MPRAGE, and DIR images. The images were generated from the reconstructed quantitative $T_1$, $T_2$, and proton density (PD) maps with 2-min 3D-QALAS using Zero-DeepSub based on Bloch equations and extended phase graph (EPG). T2-FLAIR: T2-fluid-attenuation inversion recovery; MPRAGE: magnetization-prepared rapid gradient echo; DIR: double inversion recovery; 3D-QALAS: 3D-quantification using an interleaved Look-Locker acquisition sequence with $T_2$ preparation pulse.

Supporting Information Figure S12. Reconstructed $T_1$, $T_2$, and proton density (PD) maps with fully sampled data using conventional and subspace 3D-QALAS with different regularizations: LLR, $l_1$-wavelet, and ZS-SSL for conventional QALAS, and without regularization, locally low-rank (LLR), $l_1$-wavelet, and Zero-DeepSub for subspace QALAS. For subspace QALAS, the actual reduction factor is $R^* = 4$, which was calculated by multiplying the reduction factor with the number of subspace basis. 3D-QALAS: 3D-quantification using an interleaved Look-Locker acquisition sequence with $T_2$ preparation pulse; RMSE: root mean square error.

Supporting Information Figure S13. Reconstructed subspace coefficient images with fully sampled data using subspace 3D-QALAS with different regularizations, including without regularization, LLR, $l_1$-wavelet, and Zero-DeepSub. For subspace QALAS, the actual reduction factor is $R^* = 4$, which was calculated by multiplying the reduction factor with the number of subspace basis. The signal intensities of each subspace coefficient were normalized for visualization. 3D-QALAS: 3D-quantification using an interleaved Look-Locker acquisition sequence with $T_2$ preparation pulse.

Supporting Information Figure S14. (a) Simulated 3D-quantification using an interleaved Look-Locker acquisition sequence with $T_2$ preparation pulse (3D-QALAS) signal evolution and subspace bases calculated using a singular value decomposition method. (b) Simulated magnetic resonance fingerprinting (MRF) signal evolution and subspace bases calculated using a singular value decomposition method.

**Supporting Information Data**

Supporting Information Table S1. MRI scan parameters of 3D-QALAS and turbo-FLASH sequences used for ISMRM/NIST system phantom and *in vivo* experiments.

|  | ISMRM/NIST Phantom | *In vivo* Experiment |
|---|---|---|
| **3D-QALAS** | | |
| FOV | 192 × 162 × 160 mm$^3$ | 224 × 224 × 176 mm$^3$ |
| Matrix Size | 192 × 162 × 160 | 224 × 224 × 176 |
| BW | 340 Hz/pixel | 330 Hz/pixel |
| Echo Spacing | 5.8 ms | 5.9 ms |
| Turbo Factor | 127 | 124 |
| Inversion Delay Times | [100, 1000, 1900, 2800] ms | [100, 1000, 1900, 2800] ms |
| TR | 4.5 s | 4.5 s |
| TE | 2.29 ms | 2.36 ms |
| Acceleration | 1 | 1 / 2 |
| Scan Time | 12 min 5 s | 18 min 45 s / 9 min 25 s |
| **Turbo-FLASH B$_1^+$** | | |
| FOV | 192 × 162 mm$^2$ | 224 × 224 mm$^2$ |
| Matrix Size | 64 × 52 | 64 × 64 |
| Number of Slices | 53 | 49 |
| Slice Thickness | 3 mm | 3 mm |
| BW | 340 Hz/pixel | 330 Hz/pixel |
| TR | 12.15 s | 12.91 s |
| TE | 3.24 ms | 3.16 ms |
| Acceleration | 2 | 2 |
| Scan Time | 26 s | 28 s |

3D-QALAS: 3D-quantification using an interleaved Look-Locker acquisition sequence with $T_2$ preparation pulse; turbo-FLASH: turbo-fast low-angle shot sequence for $B_1^+$ mapping. ISMRM/NIST: International Society for Magnetic Resonance in Medicine and National Institute of Standards and Technology.

Supporting Information Table S2. MRI scan parameters of IR-FSE and SE-FSE sequences used for ISMRM/NIST system phantom experiment.

|  | **IR-FSE** | **SE-FSE** |
|---|---|---|
| FOV | 192 × 192 mm$^2$ | 192 × 192 mm$^2$ |
| Matrix Size | 192 × 192 | 192 × 192 |
| Slice Thickness | 3 mm | 3 mm |
| BW | 338 Hz/pixel | 338 Hz/pixel |
| TR | 8.11 s | 1.5 s |
| TE | 7.6 ms | 10.0 ms |
| TI | [35, 100, 150, 250, 500, 1000, 2000, 3000, 4000] ms | [10, 30, 50, 70, 90, 120, 200, 300, 400] ms |
| Turbo Factor | 18 | 80 |
| Acceleration | 2 | 3 |
| Scan Time | 2 min 36 s | 2 min 3 s |

ISMRM/NIST: International Society for Magnetic Resonance in Medicine and National Institute of Standards and Technology; IR-FSE: inversion-recovery fast-spin-echo; SE-FSE: single-echo fast-spin-echo.

Supporting Information Table S3. MRI scan parameters of $T_1$w, $T_2$w, $T_2$-FLAIR, MPRAGE, and DIR sequence used for *in vivo* experiments.

|  | $T_1$w | $T_2$w | $T_2$-FLAIR | MPRAGE | DIR |
|---|---|---|---|---|---|
| FOV | 224 × 180 mm$^2$ | 256 × 176 mm$^2$ | 256 × 256 × 176 mm$^3$ | 224 × 224 × 176 mm$^3$ | 256 × 256 × 176 mm$^3$ |
| Matrix Size | 224 × 180 | 256 × 176 | 256 × 256 × 176 | 224 × 224 × 176 | 256 × 256 × 176 |
| Number of Slices | 52 | 52 | - | - | - |
| Slice Thickness | 3 mm | 3 mm | - | - | - |
| BW | 330 Hz/pixel | 331 Hz/pixel | 751 Hz/pixel | 200 Hz/pixel | 331 Hz/pixel |
| TR | 350 ms | 5000 ms | 5000 ms | 2500 ms | 7500 ms |
| TE | 3.33 ms | 87.0 ms | 391 ms | 3.37 ms | 318 ms |
| Echo Spacing | - | 7.94 ms | 3.46 ms | 7.9 ms | 5.04 ms |
| TI | - | - | 1800 ms | 1100 ms | [3000, 450 ms] |
| Flip Angle | 70 deg | 150 deg | $T_2$ VFA* | 7 deg | $T_2$ VFA* |
| Turbo Factor | - | 11 | 278 | 178 | 256 |
| Acceleration | 1 | 1 | 3 | 3 | 3 |
| Scan Time | 1 min 5 s | 1 min 27 s | 4 min 47 s | 3 min 22 s | 7 min 39 s |

* $T_2$-FLAIR and DIR sequences use $T_2$ variable flip angle (VFA) trains.

$T_2$-FLAIR: $T_2$-fluid-attenuation inversion recovery; MPRAGE: magnetization-prepared rapid gradient echo; DIR: double inversion recovery.

Supporting Information Table S4. P-values calculated from $T_1$ and $T_2$ bias comparison between conventional 3D-QALAS and subspace QALAS used in ISMRM/NIST system phantom experiment.

| P-value | (Ref. - Subspace QALAS) / Ref. (%) | | | |
|---|---|---|---|---|
| | w/o Reg. | LLR | $l_1$-wavelet | Zero-DeepSub |
| (Ref. - Conv. QALAS) / Ref. (%) | | | | |
| $T_1$ | 0.01 | 0.01 | 0.11 | 0.95 |
| $T_2$ | 0.02 | 0.47 | 0.38 | 0.02 |

* $T_1$ and $T_2$ bias of conventional and subspace QALAS were calculated from the percentage difference between the reference $T_1$ and $T_2$ values and those of conventional and subspace QALAS, respectively. Paired Wilcoxon signed-rank tests were used to calculate p-values.

3D-QALAS: 3D-quantification using an interleaved Look-Locker acquisition sequence with $T_2$ preparation pulse; ISMRM/NIST: International Society for Magnetic Resonance in Medicine and National Institute of Standards and Technology.

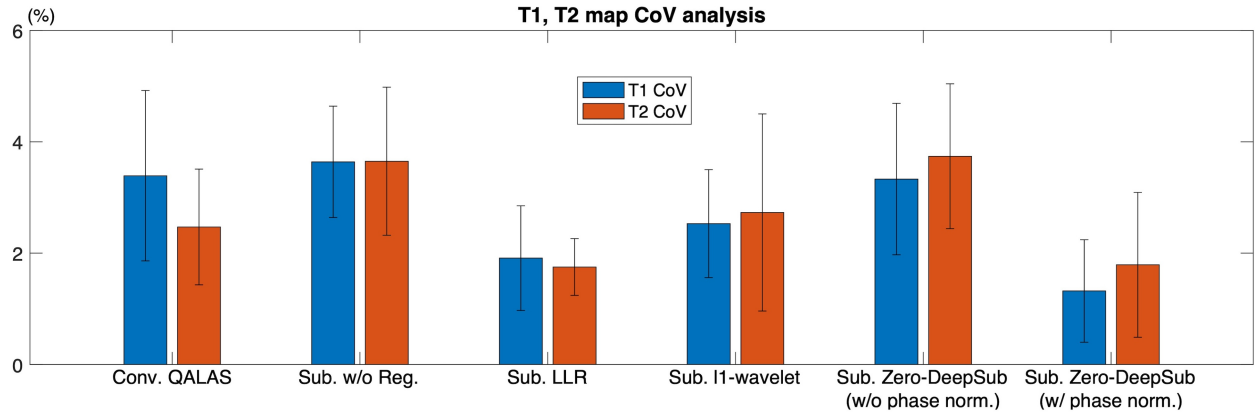

Supporting Information Figure S1. Coefficient of variation (CoV) (%) analysis of (a) $T_1$ and (b) $T_2$ values using an ISMRM/NIST system phantom with 3 × 3 reduction factor reconstructed using conventional and subspace 3D-QALAS with different regularizations, including without regularization, locally low-rank (LLR), $l_1$-wavelet, Zero-DeepSub without phase normalization, and Zero-DeepSub with phase normalization. ISMRM/NIST: International Society for Magnetic Resonance in Medicine and National Institute of Standards and Technology; 3D-QALAS: 3D-quantification using an interleaved Look-Locker acquisition sequence with $T_2$ preparation pulse.

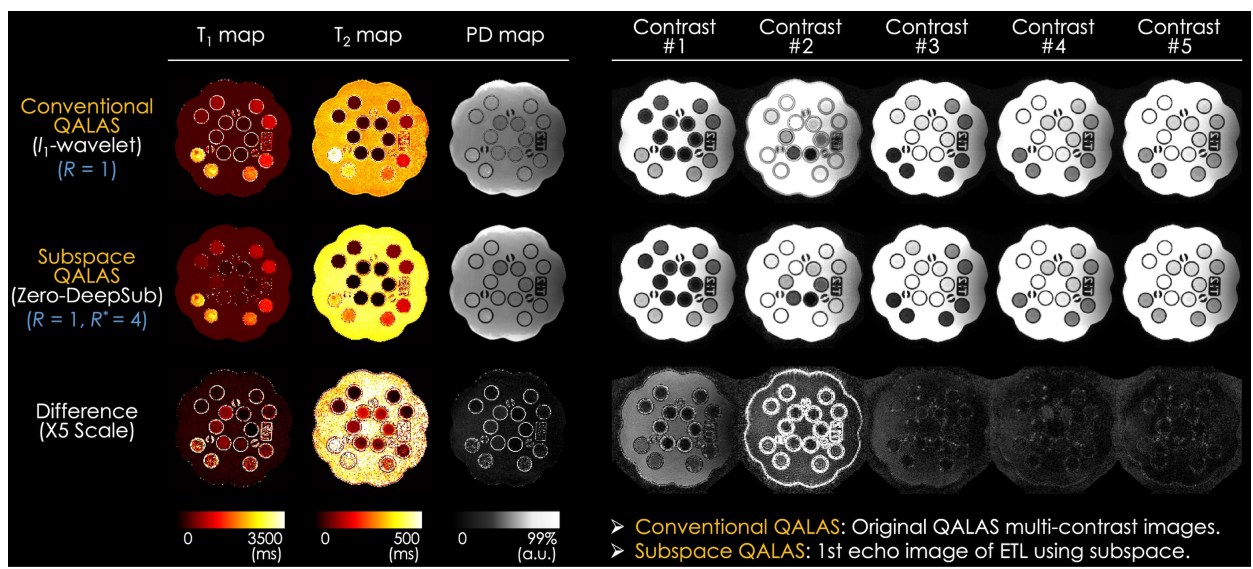

Supporting Information Figure S2. Reconstructed quantitative maps, including $T_1$, $T_2$, and proton density (PD) maps, and multi-contrast 3D-QALAS images using conventional and subspace 3D-QALAS with an ISMRM/NIST system phantom. For subspace QALAS, the actual reduction factor is $R^* = 4$, which was calculated by multiplying the reduction factor with the number of subspace basis. $l_1$-wavelet and Zero-DeepSub were used for conventional and subspace QALAS, respectively. The reconstructed multi-contrast images using conventional QALAS are original QALAS images, whereas those reconstructed using subspace QALAS are the first echo of the echo train length (ETL) multi-echo images. 3D-QALAS: 3D-quantification using an interleaved Look-Locker acquisition sequence with $T_2$ preparation pulse; ISMRM/NIST: International Society for Magnetic Resonance in Medicine and National Institute of Standards and Technology.

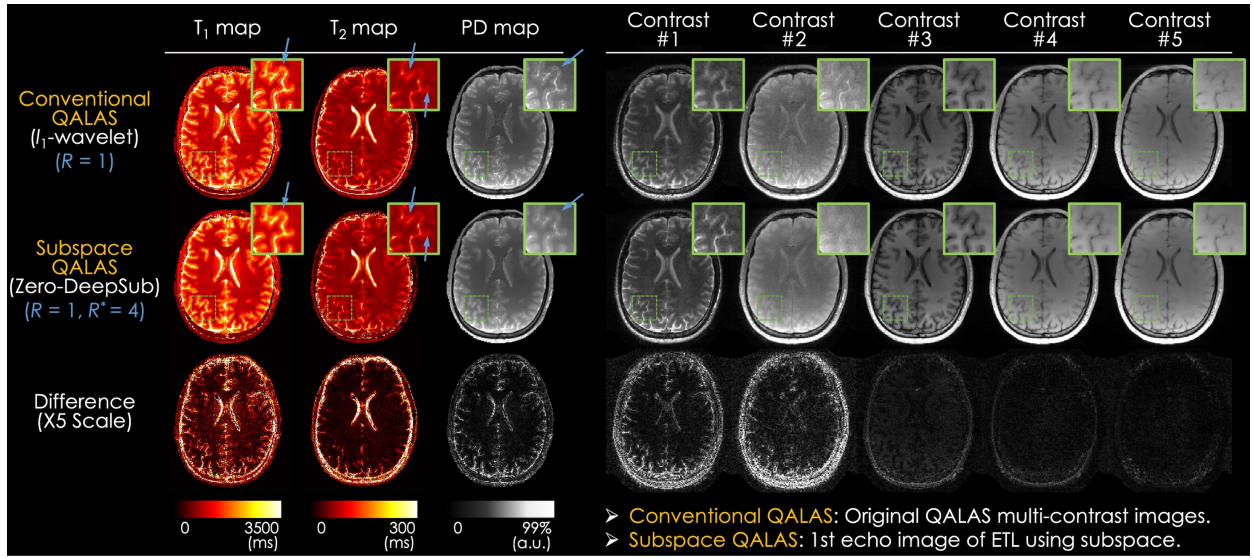

Supporting Information Figure S3. Reconstructed quantitative maps, including $T_1$, $T_2$, and proton density (PD) maps, and multi-contrast 3D-QALAS images using conventional and subspace 3D-QALAS. For subspace QALAS, the actual reduction factor is $R^* = 4$, which was calculated by multiplying the reduction factor with the number of subspace basis. $l_1$-wavelet and Zero-DeepSub were used for conventional and subspace QALAS, respectively. The reconstructed multi-contrast images using conventional QALAS are the fully sampled original QALAS images, whereas those reconstructed using subspace QALAS are the first echo of the echo train length (ETL) multi-echo images. 3D-QALAS: 3D-quantification using an interleaved Look-Locker acquisition sequence with $T_2$ preparation pulse.

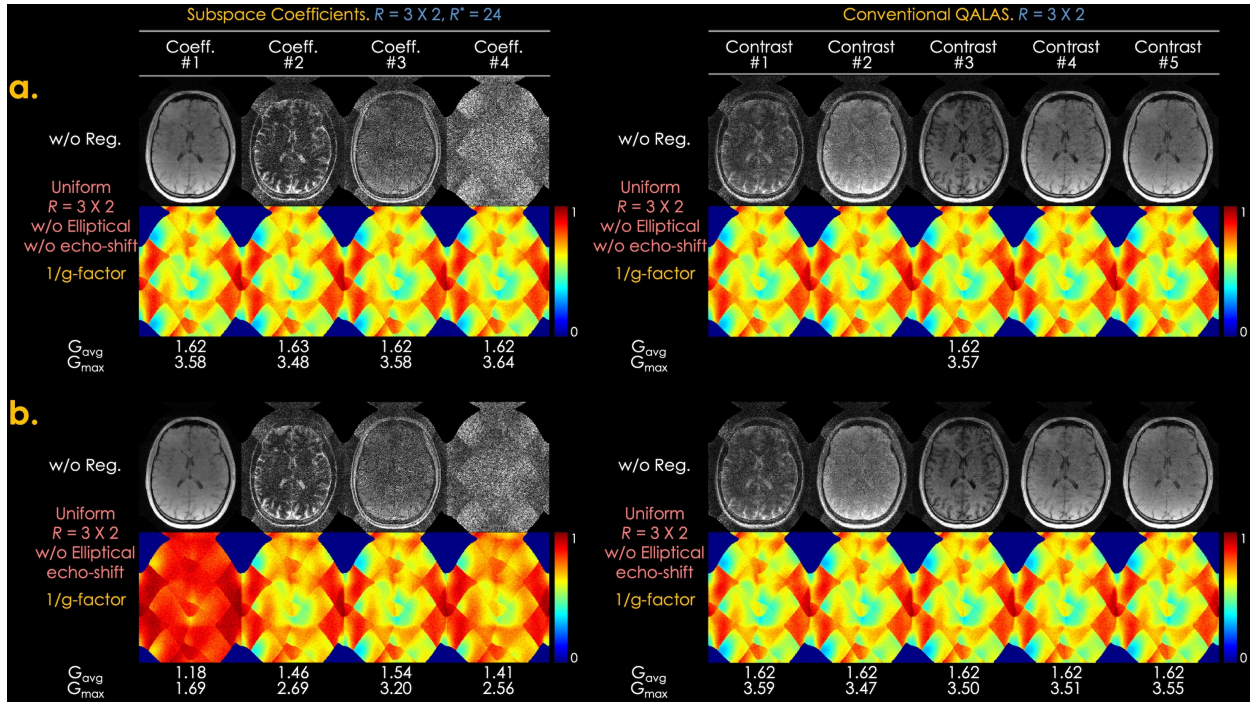

Supporting Information Figure S4. G-factor analysis of conventional and subspace 3D-QALAS with 3 × 2 reduction factor using (a) uniform without elliptical and echo-shift sampling and (b) uniform without elliptical but with echo-shift sampling patterns. For subspace QALAS, the actual reduction factor is $R^* = 24$, which was calculated by multiplying the reduction factor with the number of subspace basis. The g-factor maps of subspace coefficients were calculated for subspace QALAS, whereas individual g-factor maps of each contrast image were calculated for conventional QALAS. 3D-QALAS: 3D-quantification using an interleaved Look-Locker acquisition sequence with $T_2$ preparation pulse.

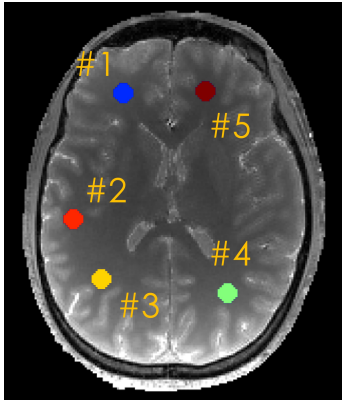

Supporting Information Figure S5. Region of interests (ROIs) used for quantitative analysis of $T_1$ and $T_2$ maps for 3D-QALAS. 3D-QALAS: 3D-quantification using an interleaved Look-Locker acquisition sequence with $T_2$ preparation pulse.

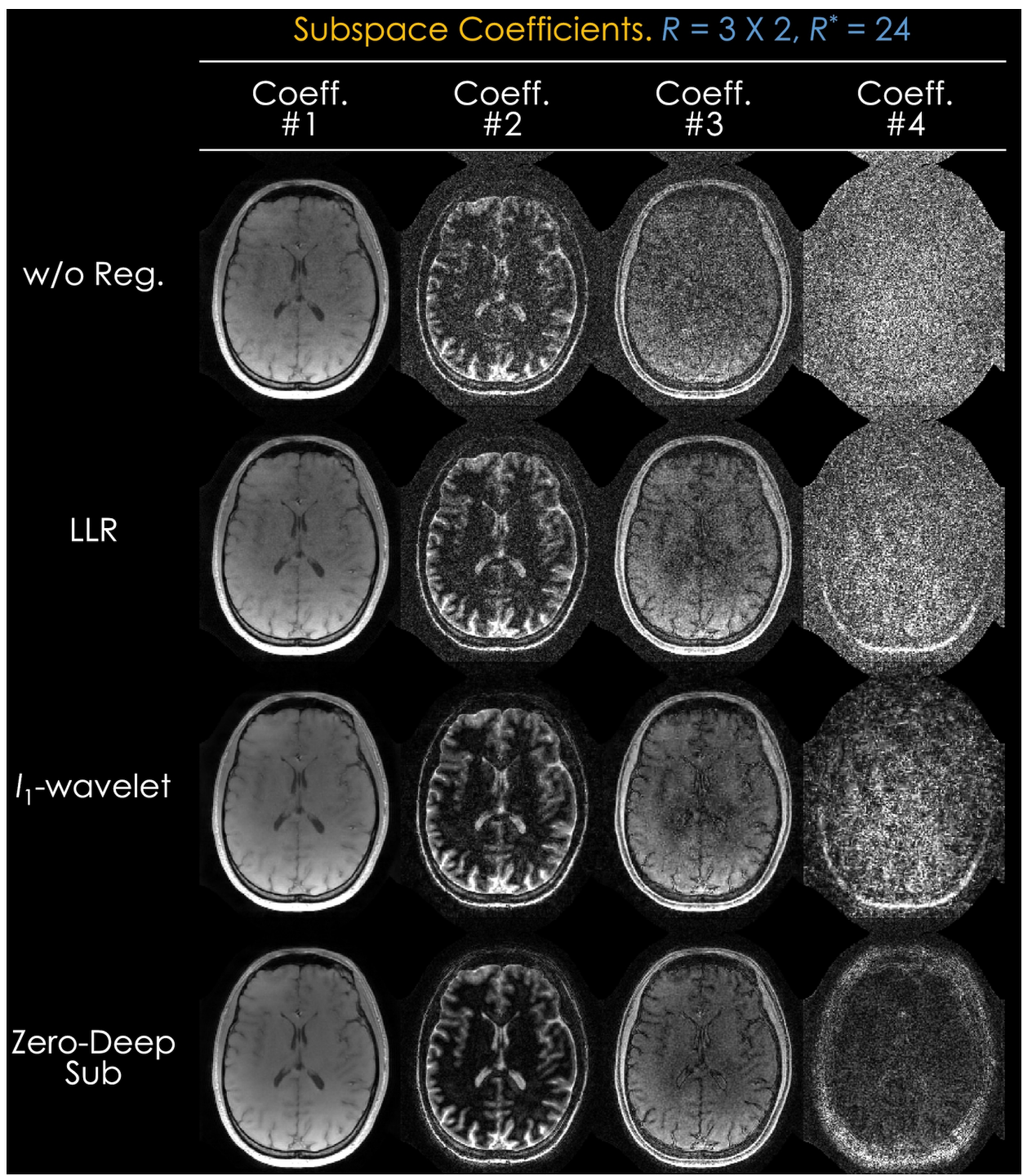

Supporting Information Figure S6. Reconstructed subspace coefficient images with Poisson sampling and 3 × 2 reduction factor using subspace 3D-QALAS with different regularizations, including without regularization, LLR, $l_1$-wavelet, and Zero-DeepSub. For subspace QALAS, the actual reduction factor is $R^* = 24$, which was calculated by multiplying the reduction factor with the number of subspace basis. The signal intensities of each subspace coefficient were normalized for visualization. 3D-QALAS: 3D-quantification using an interleaved Look-Locker acquisition sequence with $T_2$ preparation pulse.

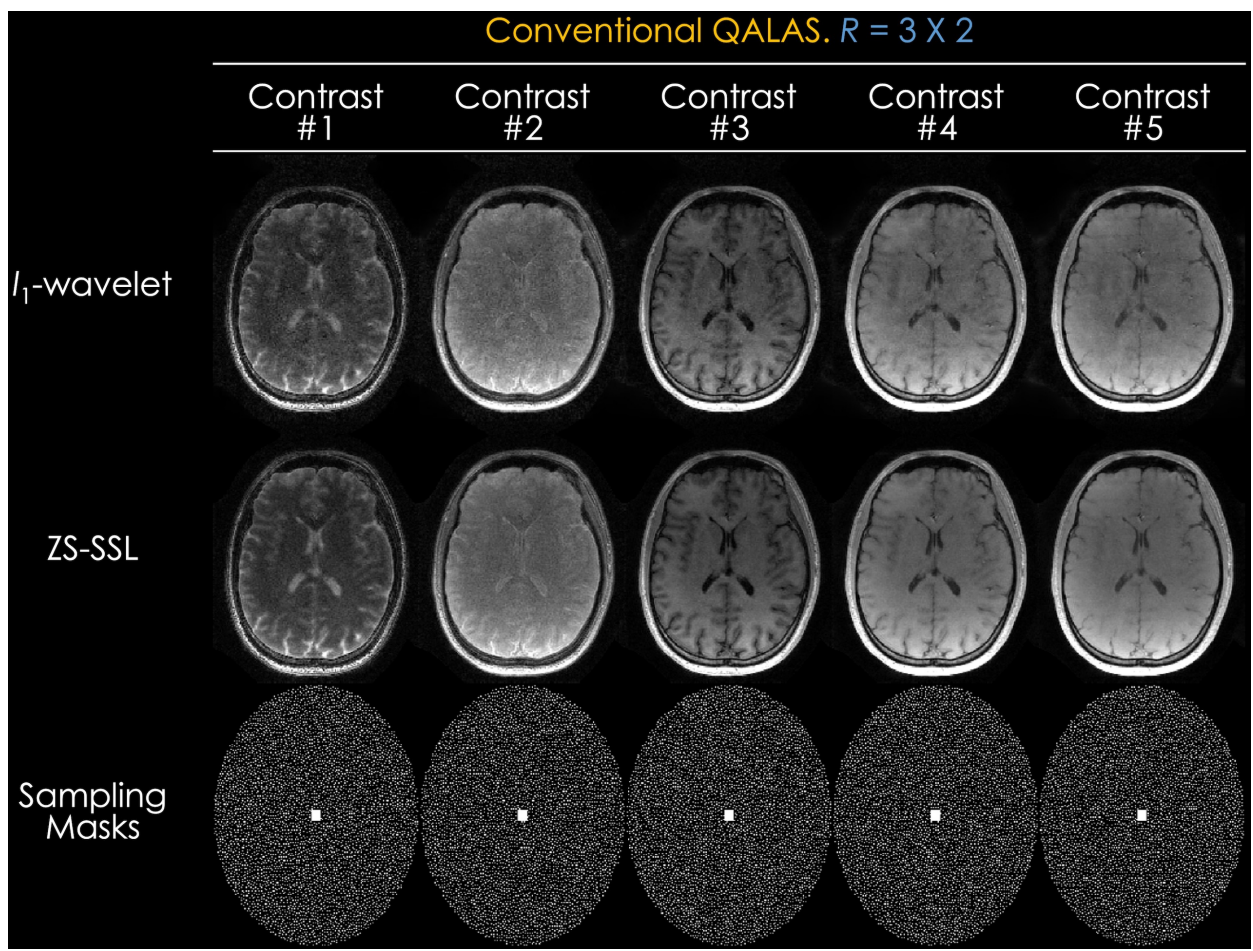

Supporting Information Figure S7. Reconstructed multi-contrast QALAS images with Poisson sampling and 3 × 2 reduction factor using conventional 3D-QALAS with different regularizations, including $l_1$-wavelet and ZS-SSL, along with the sampling masks. 3D-QALAS: 3D-quantification using an interleaved Look-Locker acquisition sequence with $T_2$ preparation pulse.

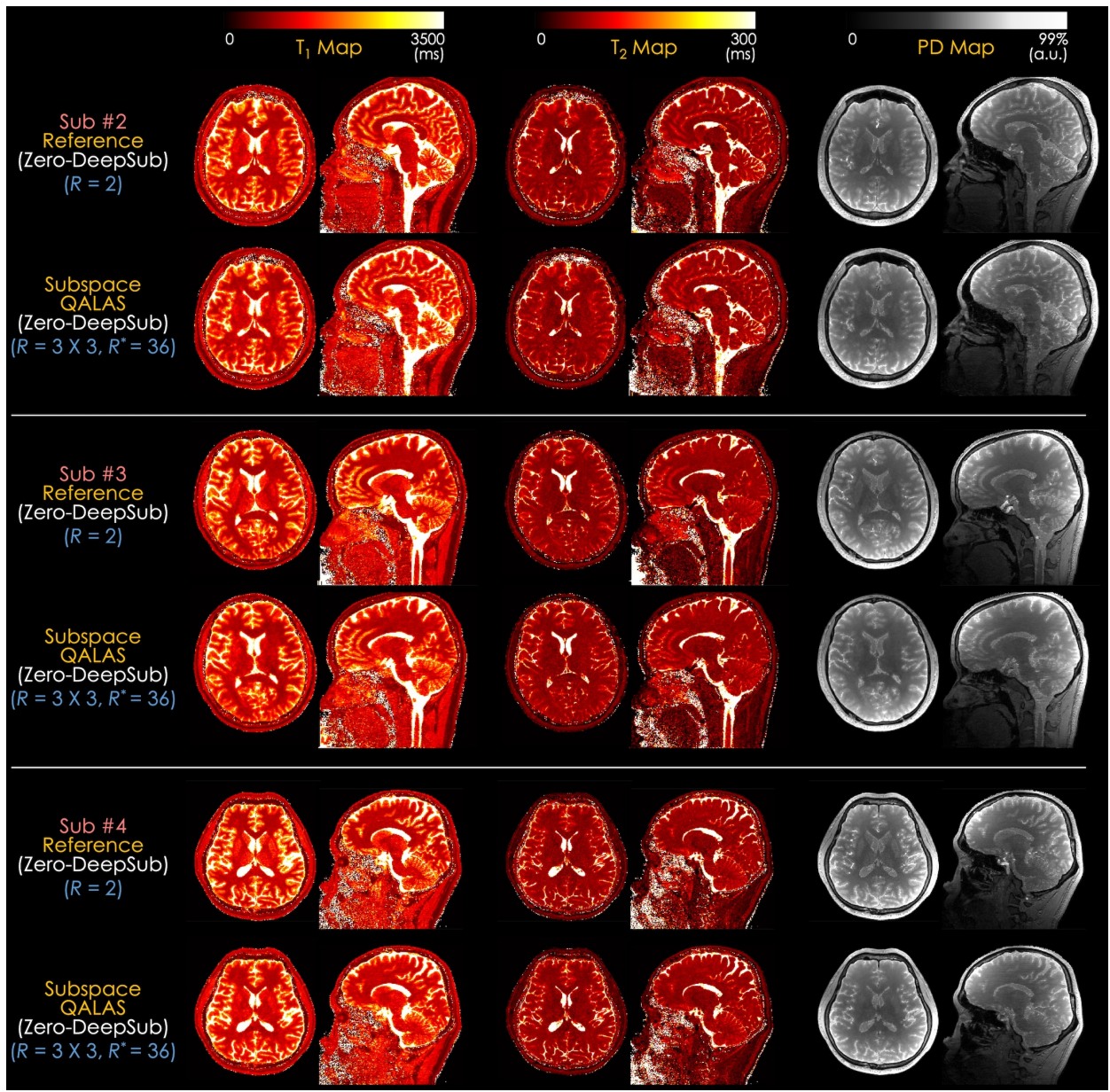

Supporting Information Figure S8. Reconstructed $T_1$, $T_2$, and proton density (PD) maps with 3 × 3 reduction factor using Poisson sampling, which enables 2 min 5 s scan time for 1mm$^3$ isotropic resolution, using subspace 3D-QALAS with Zero-DeepSub. The data were acquired with acceleration factor 2, and additional retrospective undersampling was conducted based on the Poisson sampling pattern to get 3 × 3 reduction factor. 3D-QALAS: 3D-quantification using an interleaved Look-Locker acquisition sequence with $T_2$ preparation pulse.

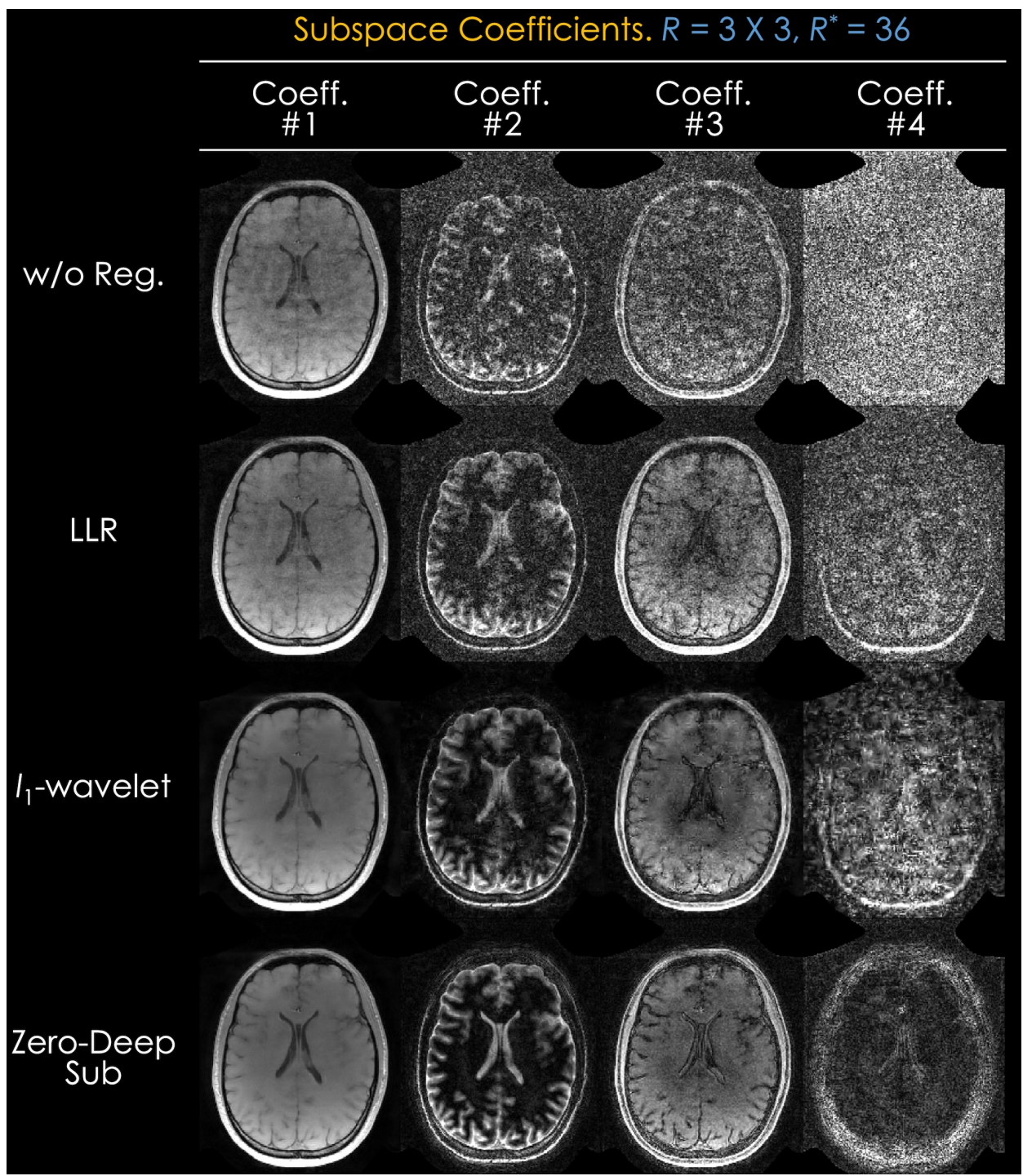

Supporting Information Figure S9. Reconstructed subspace coefficient images with Poisson sampling and 3 × 3 reduction factor using subspace 3D-QALAS with different regularizations, including without regularization, LLR, $l_1$-wavelet, and Zero-DeepSub. For subspace QALAS, the actual reduction factor is $R^* = 36$, which was calculated by multiplying the reduction factor with the number of subspace basis. The signal intensities of each subspace coefficient were normalized for visualization. 3D-QALAS: 3D-quantification using an interleaved Look-Locker acquisition sequence with $T_2$ preparation pulse.

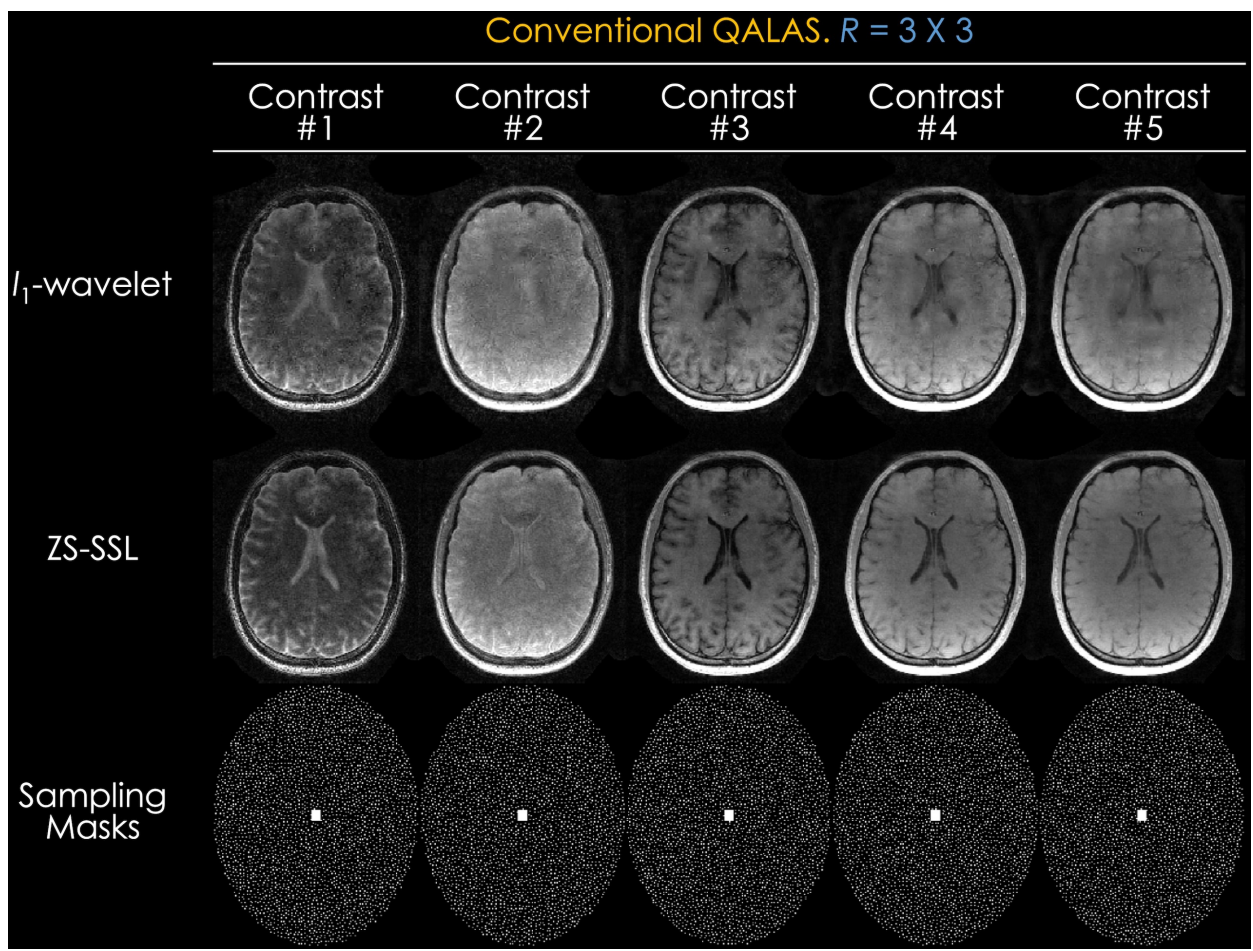

Supporting Information Figure S10. Reconstructed multi-contrast QALAS images with Poisson sampling and 3 × 3 reduction factor using conventional 3D-QALAS with different regularizations, including $l_1$-wavelet and ZS-SSL, along with the sampling masks. 3D-QALAS: 3D-quantification using an interleaved Look-Locker acquisition sequence with $T_2$ preparation pulse.

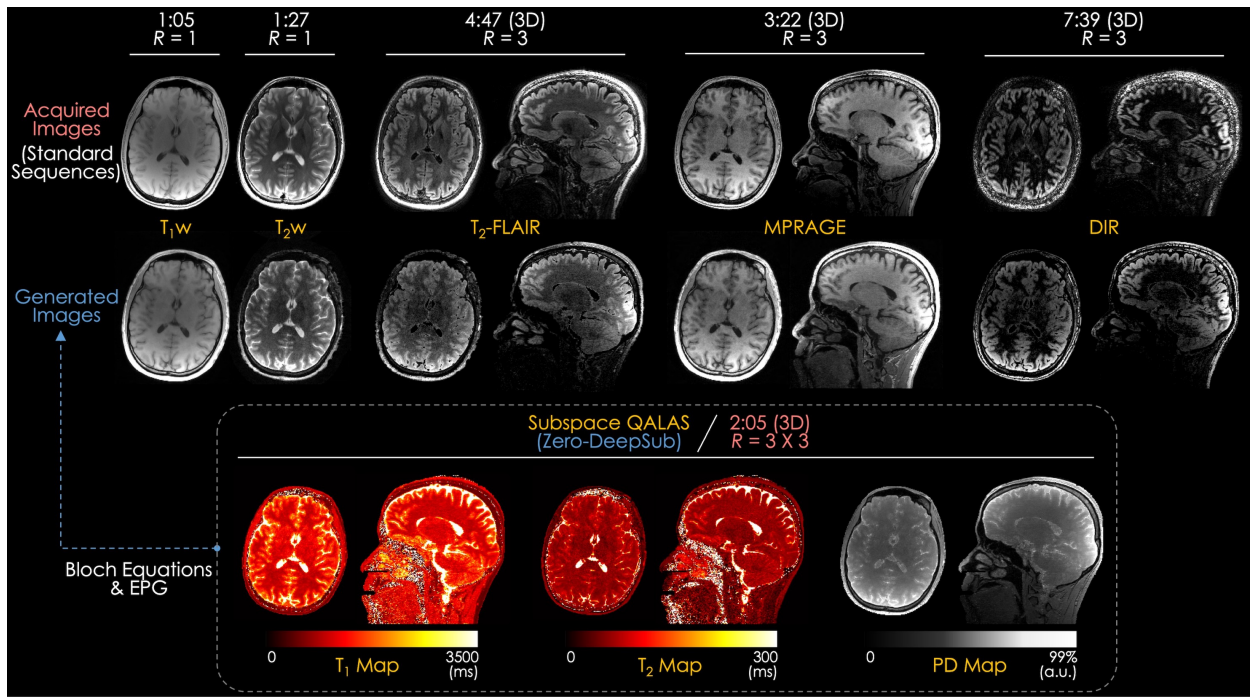

Supporting Information Figure S11. Acquired and generated $T_1$w, $T_2$w, $T_2$-FLAIR, MPRAGE, and DIR images. The images were generated from the reconstructed quantitative $T_1$, $T_2$, and proton density (PD) maps with 2-min 3D-QALAS using Zero-DeepSub based on Bloch equations and extended phase graph (EPG). T2-FLAIR: T2-fluid-attenuation inversion recovery; MPRAGE: magnetization-prepared rapid gradient echo; DIR: double inversion recovery; 3D-QALAS: 3D-quantification using an interleaved Look-Locker acquisition sequence with $T_2$ preparation pulse.

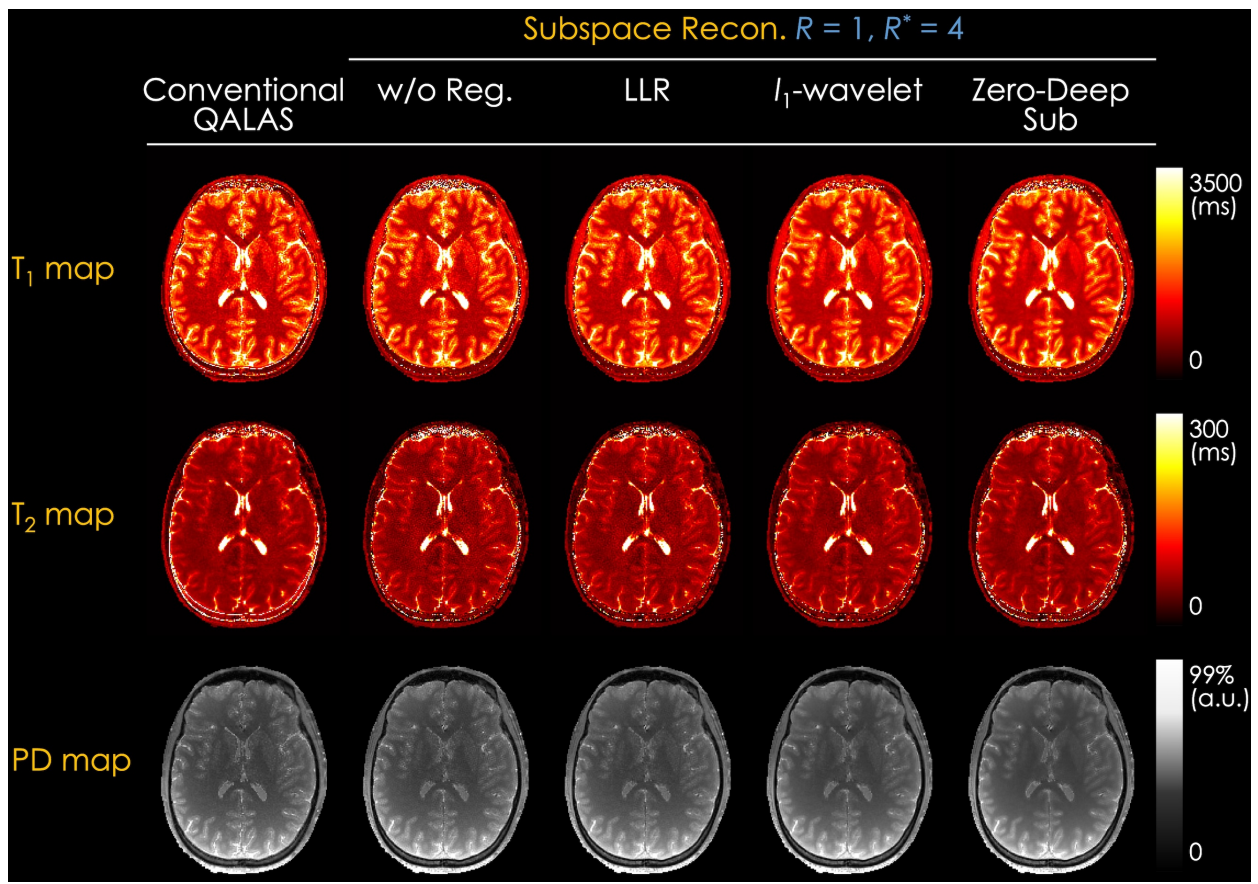

Supporting Information Figure S12. Reconstructed $T_1$, $T_2$, and proton density (PD) maps with fully sampled data using conventional and subspace 3D-QALAS with different regularizations: LLR, $l_1$-wavelet, and ZS-SSL for conventional QALAS, and without regularization, locally low-rank (LLR), $l_1$-wavelet, and Zero-DeepSub for subspace QALAS. For subspace QALAS, the actual reduction factor is $R^* = 4$, which was calculated by multiplying the reduction factor with the number of subspace basis. 3D-QALAS: 3D-quantification using an interleaved Look-Locker acquisition sequence with $T_2$ preparation pulse.

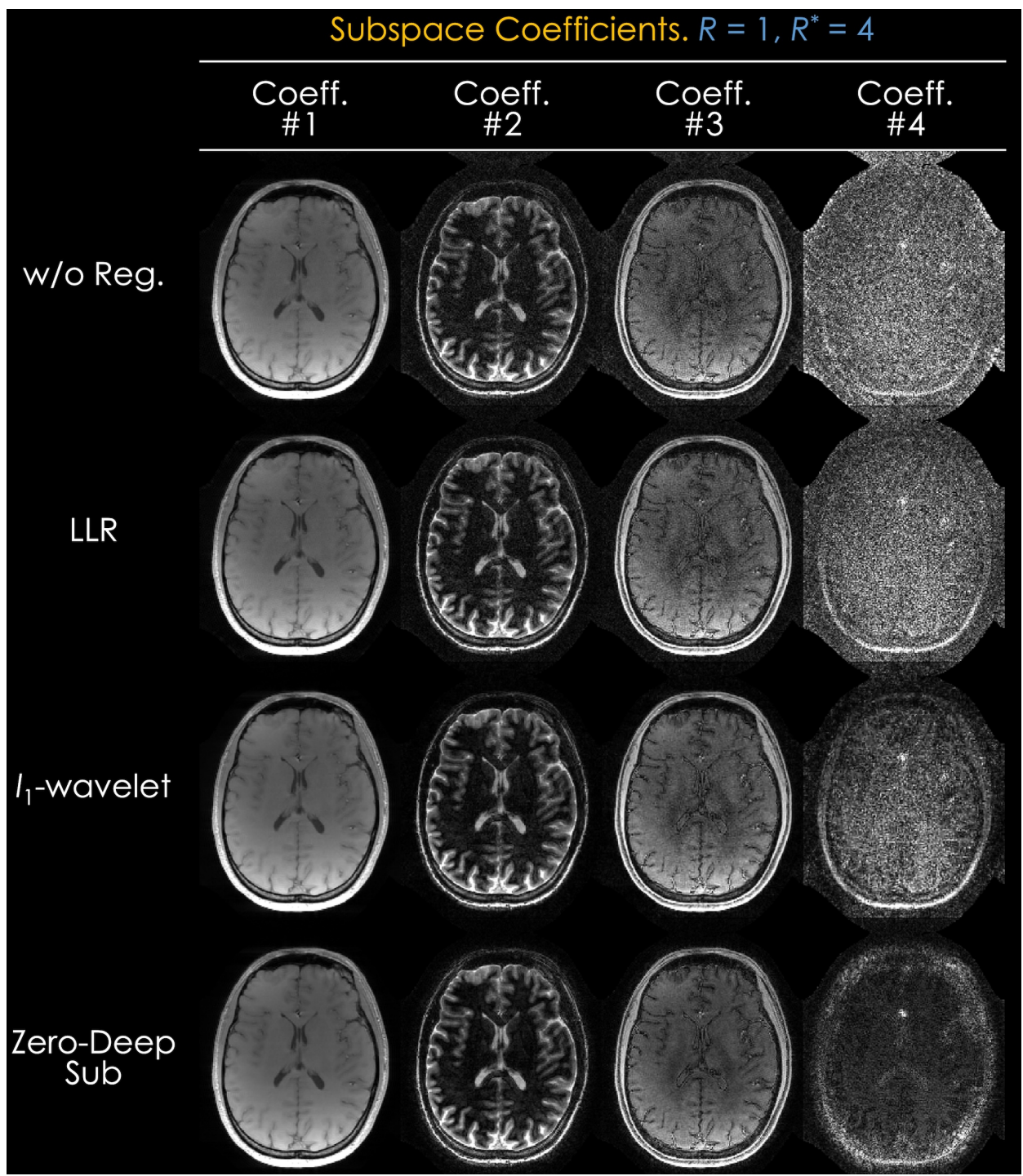

Supporting Information Figure S13. Reconstructed subspace coefficient images with fully sampled data using subspace 3D-QALAS with different regularizations, including without regularization, LLR, $l_1$-wavelet, and Zero-DeepSub. For subspace QALAS, the actual reduction factor is $R^* = 4$, which was calculated by multiplying the reduction factor with the number of subspace basis. The signal intensities of each subspace coefficient were normalized for visualization. 3D-QALAS: 3D-quantification using an interleaved Look-Locker acquisition sequence with $T_2$ preparation pulse.

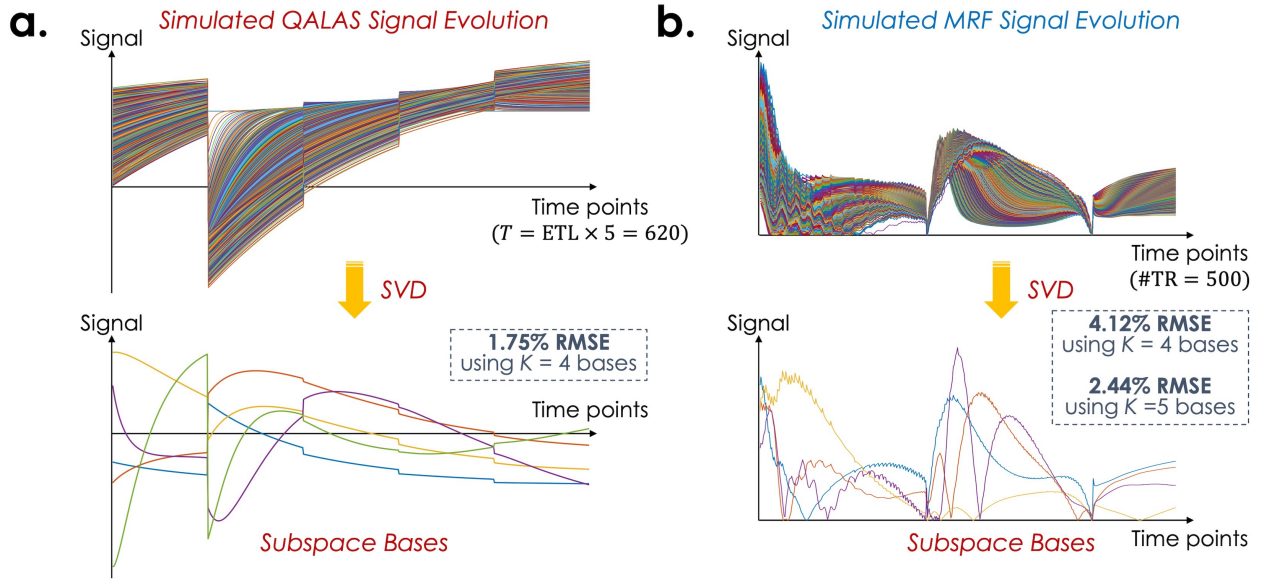

Supporting Information Figure S14. (a) Simulated 3D-quantification using an interleaved Look-Locker acquisition sequence with $T_2$ preparation pulse (3D-QALAS) signal evolution and subspace bases calculated using a singular value decomposition method. (b) Simulated magnetic resonance fingerprinting (MRF) signal evolution and subspace bases calculated using a singular value decomposition method.